\documentclass[preprint,nofootinbib,showpacs]{revtex4}
\usepackage{amsmath,amssymb,graphicx}

\begin{document}
\newcommand{\ECT}{European Centre for Theoretical Studies in Nuclear Physics
and Related Areas (E.C.T.*)\\
286 Strada delle Tabarelle, Villazzano (Trento), I-38050,~Italy.}
\newcommand{\be}{\begin{eqnarray}}
\newcommand{\ee}{\end{eqnarray}}
\newcommand{\GeV}{\textrm{GeV}}
\newcommand{\fm}{\textrm{fm}}
\newcommand{\la}{\langle}
\newcommand{\ra}{\rangle}
\newcommand{\Tr}{\textrm{Tr}}

\bibliographystyle{apsrev}

\title{Instanton Contribution to the 
Electro-Magnetic Form Factors of the Nucleon}
\author{P. Faccioli}
\affiliation{\ECT\\and\\ I.N.F.N., Gruppo Collegato di Trento.}


\begin{abstract}

We study the electro-magnetic form 
factors of the nucleon, from small to large momentum transfer, in the
context of the Instanton Liquid Model (ILM).
As a first step, we analyze the role of single-instanton effects, and
show that they dominate the form factors at large momentum transfer.
Then, we go beyond the single-instanton approximation and perform
a calculation to all order in the 't~Hooft interaction.
We find that the ILM is in good agreement 
with the available experimental data.
Based on these results, we argue that instantons 
provide a microscopic mechanism that explains the delay of the 
onset of the asymptotic perturbative regime, in the 
electro-magnetic form factors.

\end{abstract}


\maketitle

\section{Introduction}
\label{introduction}
The recent measurements of pion and nucleon form factors
performed at Jefferson Laboratory (JLAB) have triggered an important 
discussion about the transition from the non-perturbative 
to the perturbative regime, in QCD.
The pion form factor has been measured very accurately from 
$0.6~\textrm{GeV}^2<Q^2<1.6~\textrm{GeV}^2$ by the $F_\pi$ Collaboration
\cite{Fpi}. It was found that the form factor 
deviates significantly from 
the asymptotic perturbative prediction, even at the 
largest value of the momentum transfer. 
Important information about the proton form factors 
has been obtained by means of the recoil polarization method, which allows 
to access the ratio of the electric over magnetic form factors 
\cite{JLAB1,JLAB2}.
These experiments have shown that the ratio $\mu\,G_E(Q^2)/G_M(Q^2)$
decreases very rapidly, while in the asymptotic regime it
should approach constant \cite{BF}.

These two results have indicated that, in elastic form factors,
the asymptotic perturbative regime is not reached until very large
values of the momentum transfer. 
Interestingly, this conclusion  contrasts with the results of the CLEO 
experiment on $\gamma\,\gamma^*\to \pi_0$ transition form factor, where
the asymptotic regime is reached already at $Q^2\gtrsim~2~\textrm{GeV}^2$.
For completeness, it should be mentioned that
there exists also a combination of proton form factors which seems
to exhibit a precocious scaling toward the perturbative behavior, 
namely the ratio of
Pauli over Dirac form factors $F_2(Q^2)/F_1(Q^2)$, where:
\be
F_1(Q^2)&:=&\frac{1}{1+\tau}\,\left(G_E(Q^2)+\tau\,G_M(Q^2)\right)\\
\label{F1def}
F_2(Q^2)&:=&\frac{1}{1+\tau}\,\left(G_M(Q^2)- G_E(Q^2)\right), \qquad 
\tau:=\frac{Q^2}{4\,M^2}.
\label{F2def}
\ee
It was recently shown by Belitsky, Ji and Yuang that, when logarithmic 
corrections and sub-leading twist light-cone wave-functions are introduced, 
perturbative QCD predicts \cite{Ji}:
\be
\label{pQCDF2F1}
\frac{F_2(Q^2)}{F_1(Q^2)}\sim \,\log^2(Q^2\Lambda^2_{QCD})/Q^2,
\ee
in good agreement with experimental data.
On the other hand, the perturbative predictions
for the individual Pauli and the Dirac form factor are nevertheless
very far from the experimental data.
This fact lead the authors to argue  that the precocious scaling
of the ratio (\ref{pQCDF2F1}) could be the result of a delicate
cancellation in the numerator and denominator.

The delay of the onset of the perturbative regime in the elastic form factors
implies that 
there are strong 
non-perturbative forces inside hadrons, which  
dominate over the perturbative gluon-exchange 
even at short distances (of the order of $1/Q$).
Two important theoretical questions arise from this fact.
What is the microscopic
origin of the short-scale interaction, driving 
the pion and nucleon elastic form factors away from their perturbative 
limit?
Why such non-perturbative
forces do not show up in the $\gamma\,\gamma^*\to \pi_0$ transition
form factor? 

Clearly, the answers to these questions 
reside in the non-perturbative sector of QCD.
In particular, it is commonly accepted that the soft physics  
light quarks is very much influenced by the interactions
responsible for  chiral symmetry breaking (CSB).
On the other hand, confinement seems to play only a marginal role.
The most convincing evidence in this direction comes from lattice studies of
 QCD in the semi-classical limit: by means of the so-called
 ``cooling'' procedure, 
it was observed that, in this limit, 
the current-current correlators of light hadrons change very little, although 
all perturbative fluctuations are removed, and the string tension 
drops out\cite{chu94}.

The characteristic scale 
associated to CSB is  $4\,\pi f_\pi \sim 1.2~\textrm{GeV}$, 
significantly larger than the typical confinement scale, 
$\Lambda_{QCD}$.
Such a separation  justifies attempting understand the 
short-distance non-perturbative structure of light-hadrons, 
without having simultaneously to account for the
 microscopic origin of confinement. 
On the other hand, from the observation that $4\,\pi\,f_\pi \sim m_{\eta'}$
it follows that any effective 
description of the short-distance 
non-perturbative dynamics of light quarks 
should also account for topological effects.

Instantons are topological 
gauge configurations which dominate the QCD Path Integral 
in the semi-classical limit. They generate the so-called 't~Hooft interaction,
that solves 
the U(1) problem\cite{'thooft} and spontaneously breaks 
chiral symmetry \cite{dyakonov}, but does not confine. 
Evidence for such an instanton-induced interaction in QCD 
comes from a number of  
phenomenological studies \cite{shuryakrev}, as well as from lattice simulations
\cite{chu94, Degrand, scalar}.
The ILM assumes that the QCD vacuum is saturated by an
ensemble of instantons and anti-instantons. 
The only phenomenological parameters in the model are the average instanton
size  $\bar{\rho}\simeq~1/3$~fm and density 
$\bar{n}\simeq~1~\textrm{fm}^{-4}$. These values
were extracted  more than two decades ago, from the global 
vacuum properties~\cite{shuryak82}.
The non-perturbative contribution to the electro-magnetic form factors of
 the nucleon  has been analyzed in a number of works by means of  
phenomenological models (for an incomplete list see \cite{incomplete}).

It the present study, we use the ILM to address the question whether
the 't~Hooft interaction can
provide the non-perturbative dynamics needed to explain the 
experimental results on pion and nucleons's form factors.
The instanton contribution to these form factors has been investigated 
in the context of the ILM in a number of works.
In~\cite{forkel} Forkel and Nielsen computed
the pion form factor in a sum-rule approach, which takes into account 
the direct-instanton contribution, in addition to the lowest dimensional
condensate terms in the 
operator product expansion\footnote{This calculation
has been recently repeated, including both  NLO perturbative 
corrections and a more realistic estimate of the
single-instanton contribution \cite{forkel2}}.
As in other sum-rule approaches, this calculation 
required a detailed knowledge of the contribution coming from 
the continuum of excitations. 
In  order to avoid this problem, in \cite{blotz} and \cite{3ptILM} the
electro-magnetic pion and proton three-point 
functions were calculated in coordinate space, by means of numerical
simulations in the ILM. The contribution from the 
continuum of excitations could be excluded by considering sufficiently 
large-sized correlation functions.
The results were then compared to phenomenological estimates of the 
same correlation functions, obtained from the Fourier 
transform  of the fits of the experimental data. 
Unfortunately, this method has the 
shortcoming that it does not allow a {\it direct} comparison of the 
theoretical predictions against the experimental data.

Direct comparison  between theory and form factors at intermediate and
large momentum transfer became possible, after the Single Instanton
Approximation (SIA) was developed \cite{sia, mymasses}.
In \cite{pionFF} it was shown that instantons can quantitatively
explain the pion charged form factor and its deviation 
from the perturbative regime at large momentum transfer. Conversely, 
it was observed that such effects are parametrically  suppressed 
in the $\gamma\, \gamma^*\to \pi_0$ transition form factor.
This explains the early onset of the perturbative regime in such a form factor.
Moreover, a calculation of the
pion distribution amplitude in the ILM 
was performed in \cite{Dorokhov}.
It was found that instantons can explain the behavior 
of the low-energy experimental data ($Q^2<~2$~GeV$^2$) 
for the  $\gamma\,\gamma^\star~\to~\pi^0$ transition form factor.

The single-instanton contribution to the nucleon electric 
from factors were first investigated in \cite{nucleonGE}. 
In this work we extend the analysis to the magnetic as well as to the
 Pauli and Dirac form factors of the nucleon.
Moreover, we also go beyond the single-instanton approximation
and include many-instanton effects, by performing
a calculation to all orders in the 't~Hooft interaction.
We shall find that experimental data at large momentum transfer can be
reproduced surprising well in the SIA. On the other hand,
form factors at low momenta are very sensitive to many-instanton effects.
In general, we have found very good agreement between theory and 
experiment, which
indicates that instantons provide the correct  
non-perturbative dynamics, 
 responsible for the electro-magnetic structure of the nucleon and for
the delay of the onset of the 
perturbative asymptotic regime in elastic form factors.

The paper is organized as follows. In section \ref{Euclidean} we review the
connection between the form factors of the nucleon and some 
Euclidean correlation
functions, which have to be evaluated non-perturbatively.
In section \ref{siaFF} we introduce the SIA and present the predictions
for the Sachs as well as for the Dirac and Pauli form factors of the nucleon.
In section \ref{MIGE} we include many-instanton effects by means of
numerical simulations in  the full-instanton liquid.
All results are summarized in section \ref{conclusions}, while the appendix
contains a compilation of the analytic SIA results.

\section{Form Factors and Euclidean Correlation Functions}
\label{Euclidean}
In order to compute the form factors of the nucleon we consider the following
Euclidean correlation functions:
\be
{\bf G3}^{p(n)}_E(t,{\bf q}) =
\int d^3{\bf x} \, d^3{\bf y} \, e^{i \, {\bf q} \cdot ({\bf x} +
{\bf y})/2} \, \langle 0 | \, \text{Tr} \,
[\,\eta_{\text{sc}}^{p(n)}(t,{\bf y}) \, J_4^{em}(0,{\bf 0}) \,
\bar{\eta}_{\text{sc}}^{p(n)}(-t,{\bf x}) \, \gamma_4 \,] | 0 \rangle,
\label{G3E}\\
{\bf G3}^{p(n)}_M(t,{\bf q}) = \int d^3{\bf x} \,
d^3{\bf y} \, e^{i \, {\bf q} \cdot ({\bf x}+{\bf y})/2}
\langle 0 | \, \textrm{Tr} [ \, \eta_{\textrm{sc}}^{p(n)}
(t,{\bf y}) \, J_2^{em}(0,{\bf 0}) \,
\bar{\eta}_{\textrm{sc}}^{p(n)}(-t,{\bf x}) \,\gamma_2 \, ] \, | 0 \rangle,
\label{G3M}
\ee
where $J_\mu^{em}(x)$ is the electro-magnetic current
and $\eta^{p(n)}_{\textrm{sc}}(x)$ is an operator 
which excites states with the quantum numbers of the nucleon. 
In the case of the proton we choose\footnote{The corresponding operator for the
neutron is obtained through the substitution $u \leftrightarrow d$.}:
\be
\label{current}
\eta^{p}_{\textrm{sc}}(x) = \epsilon^{a b c} \, [ u^T_a(x) C \gamma_5 d_b(x) ]
\, u_c(x).
\ee
In QCD, in the limit of large Euclidean time separation $t$, 
the correlation functions
(\ref{G3E}) and (\ref{G3M}) relate directly to the form factors of the
nucleon. In particular ${\bf G3}^{p(n)}_E$ depends linearly on the proton
 (neutron) electric form factor:
\be
\label{spectralG3E}
{\bf G3}^{p(n)}_E(t,{\bf q}) 
 &\rightarrow& 8\,M^2 \,R(t,{\bf q})\, G^{p(n)}_{\text{E}}(Q^2), \\
R(t,{\bf q}) &:=& \Lambda^2_{\text{sc}}\,
\left(\frac{1}{2\,\omega_{{\bf q}/2}}\right)^2
 \,e^{-2 \,\omega_{{\bf q}/2} \, t},
\ee
where $G^{p(n)}_{\text{E}}(Q^2)$ denotes the proton (neutron)
electric form factor and $\Lambda_{\textrm{sc}}$ the coupling of the interpolating
operator (\ref{current}) to the nucleon. 
Similarly, if ${\bf q}$ is chosen along the $\hat{x}$ direction,
 ${\bf G3}^{p(n)}_M$
relates to the proton (neutron) magnetic form factor:
\be
\label{spectralG3M}
{\bf G3}^{p(n)}_M(t,{\bf q}) \rightarrow -2\,{\bf q}^2 
\,R(t,{\bf q})\, G^{p(n)}_{\text{M}}(Q^2).
\ee
These expressions are derived in the Breit-Frame, where 
${\bf p'}=-{\bf p}={\bf q}/2$ and $Q^2={\bf q}^2$.
We recall that, in the  kinematic regime explored by current experiments,
both Sachs form factors are positive definite.
This implies that ${\bf G3}^{p(n)}_M(t,{\bf q})$ and 
${\bf G3}^{p(n)}_E(t,{\bf q})$ have opposite sign.

From the correlation functions (\ref{G3E}) and (\ref{G3M}) 
it is immediate to construct
linear combinations which relate to the Dirac and Pauli form factors:
\be
{\bf G3}^{p(n)}_{F_1}
:&=&\qquad {\bf G3}^{p(n)}_{E} \,-{\bf G3}^{p(n)}_M \rightarrow
 8\,M^2\,(1+\tau)\, R(t,{\bf q})\,F_1^{p(n)}(Q^2)
\label{spectralG3F1}\\
{\bf G3}^{p(n)}_{F_2}
:&=&-\frac{1}{\tau}\;{\bf G3}^{p(n)}_{M} \,-{\bf G3}^{p(n)}_E\, \rightarrow
 8\,M^2\,(1+\tau)\, R(t,{\bf q})\,F_2^{p(n)}(Q^2)
\label{spectralG3F2}.
\ee
Notice that, due to the sign difference, the absolute contributions of  
the correlation functions ${\bf G3}_M$ and ${\bf G3}_E$ 
to the Dirac form factor
 $F_1(Q^2)$ are added-up, while the contributions to the Pauli form factor
$F_2(Q^2)$ are subtracted.

The exponential factor $R(t,{\bf q})$ 
in (\ref{spectralG3E})-(\ref{spectralG3M}) and 
(\ref{spectralG3F1})-(\ref{spectralG3F2}) can be obtained from
the two-point function:
\be
\label{G2}
{\bf G2}(t,{\bf q}) = \int d^3{\bf x}
\, e^{i \, {\bf q} \cdot {\bf x}} \, \langle 0 | \, \text{Tr} \,
\eta_{\text{sc}}(t,{\bf x}) \, \bar{\eta}_{\text{sc}}(0) \,
\gamma_4 \, | 0 \rangle.
\ee
In the large Euclidean time limit, one has:
\be
\label{spectralG2}
{\bf G2}(t,{\bf q}) \longrightarrow 2 \,
\Lambda^2_{\text{sc}} \, e^{-\omega_{\bf q} \, t},
\ee
from which it is possible to extract the coupling constant
$\Lambda_{\textrm{sc}}$ and the nucleon mass $M$.

Even at asymptotically large momentum transfer, 
the correlation functions defined in this section cannot be calculated in
perturbation theory.
This is because all the three-point and two-point 
functions are large-sized  (due to the $t\to\infty$ limit), 
while pQCD is supposed to work only for small-sized correlation functions.
On the other hand, factorization theorems
state that, at asymptotically large values of the momentum transfer,
all non-perturbative effects are included in the light-cone wave functions
and decouple from the hard perturbative contributions. 
The problem with such an approach is that it is not possible to know a priori
at which momenta factorization theorems become quantitatively reliable.
Therefore, in this work we shall refrain from using them
and attempt a {\it direct} non-perturbative 
evaluation of the Green functions, from moderate to large $Q^2$.

After performing Wick contractions, the fermionically connected components 
of the correlators (\ref{G3E}) and (\ref{G3M}) read:
\be
{\bf G3}_{E(M)}(t,{\bf q})=   
\int d^3{\bf x} \int d^3{\bf y} 
\, e^{i \, {\bf q}/2 \cdot ({\bf x}+{\bf y})}\,
\epsilon_{a b c}\,\epsilon_{a' b' c'}\nonumber\\
\la 
U^{a b c a' b' c'}_{A\,4(2)} +   
U^{a b c a' b' c'}_{B\,4(2)} + 
U^{a b c a' b' c'}_{C\,4(2)} + 
U^{a b c a' b' c'}_{D\,4(2)} + 
D^{a b c a' b' c'}_{A\,4(2)} + 
D^{a b c a' b' c'}_{B\,4(2)} \ra,
\label{G3wick}
\ee
where
\be
U^{a b c \,a' b' c'}_{A\,4(2)}  = 
\Tr [ 
S_{c  b'}(t,{\bf y};-t,{\bf x})
(C\,\gamma_5)^T \,
S^T_{a  a'} (t,{\bf y};-t,{\bf x})
(C\,\gamma_5) \,\nonumber\\
S_{b  e}(t,{\bf y};0,{\bf 0})
\gamma_{4(2)}
S_{e  c'}(0,{\bf 0};-t,{\bf x})
\gamma_{4(2)}]\nonumber\\
U^{a b c \,a' b' c'}_{B\,4(2)} = 
-\Tr[ 
\gamma_{4 (2)} 
S_{cc'}(t,{\bf y};-t,{\bf x})]\,
\Tr[ 
(C\,\gamma_5)\,
S_{be}(t,{\bf y};0,{\bf 0})
\gamma_{4(2)}\nonumber\\
S_{eb'}(0,{\bf 0};-t,{\bf x})
(C\,\gamma_5)^T \,
S^T_{aa'} (t,{\bf y};-t,{\bf x})]\nonumber\\
U^{a b c \,a' b' c'}_{C\,4(2)}  = 
\Tr [
S_{ce}(t,{\bf y};0,{\bf 0})
\gamma_{4(2)}
S_{eb'}(0,{\bf 0};-t,{\bf x})
(C\,\gamma_5)^T 
S_{cb'}(t,{\bf y};-t,{\bf x})
(C\,\gamma_5)^T \,\,\nonumber\\
S^T_{aa'} (t,{\bf y};-t,{\bf x})
(C\,\gamma_5)\,
S_{bc'} (t,{\bf y};-t,{\bf x})
\gamma_{4(2)}]\nonumber\\
U^{a b c \,a' b' c'}_{D\,4(2)} = 
-\Tr [
(C\,\gamma_5)\,
S_{bb'}(t,{\bf y};-t,{\bf x})
(C\,\gamma_5)^T \,
S_{aa'}^T(t,{\bf y};-t,{\bf x})]\,
\nonumber\\\Tr[S_{ce}(t,{\bf y};0,{\bf 0})
\gamma_{4(2)}
S_{ec'}(0,{\bf 0};-t,{\bf x})
\gamma_{4(2)}],\nonumber
\ee
and
\be
D^{a b c \,a' b' c'}_{A\,4(2)} = 
-\Tr[ 
\gamma_{4 (2)} 
S_{cb'}(t,{\bf y};-t,{\bf x})]\,
(C\,\gamma_5)^T\,
S^T_{aa'} (t,{\bf y};-t,{\bf x})
(C\,\gamma_5)\,\nonumber\\
S_{be}(t,{\bf y};0,{\bf 0})
\gamma_{4(2)}S_{eb'}(0,{\bf 0};-t,{\bf x})]
\nonumber\\
D^{a b c \,a' b' c'}_{B\,4(2)}  = 
\Tr [
S_{a  e}(t,{\bf y};0,{\bf 0})
\gamma_{4(2)}
S_{e  a'}(0,{\bf 0};-t,{\bf x})
\,(C\,\gamma_5) \,\,\nonumber\\
S_{c  b'}^T(t,{\bf y};-t,{\bf x})
\gamma_{4(2)^T}
S_{b  c'} (t,{\bf y};-t,{\bf x})
(C\,\gamma_5)^T]\nonumber.
\ee
Similarly, the two-point function reads:
\be
{\bf G2} (t,{\bf p}) = \int d^3{\bf x} \, 
e^{i \, {\bf p} \cdot {\bf x}}
\epsilon_{a b c}\,\epsilon_{a'b'c'}
\la~N_A^{a b c \ a'b'c'} + N_B^{a b c \ a'b'c'}~\ra, 
\label{G2wick}
\ee
where,
\be
N_A^{a b c\ a'b'c'}=(-1)~\
\Tr[~S_{a  a'} (t,{\bf x};0,{\bf 0})
\,(C\,\gamma_5) \,S^T_{b  b'} (t,{\bf x};0,{\bf 0})
(C\,\gamma_5)^T]\,\Tr[S_{c  c'} (t,{\bf x};0,{\bf 0})\,
\gamma_4~]\nonumber\\
N_B^{a b c\ a'b'c'}=~
\Tr[~S_{a  a'} (t,{\bf x};0,{\bf 0})
\,(C\,\gamma_5) \,S^T_{c  b'} (t,{\bf x};0,{\bf 0})\,\gamma_4\,
S^T_{b  c'} (t,{\bf x};0,{\bf 0})\,\,(C\,\gamma_5)^T~].\nonumber
\ee
In these expressions, $S(y_4,{\bf y}; x_4, {\bf x})$ 
denotes the quark propagator, the trace is over spinor and color indices, 
and the brackets 
$\langle \: \cdot \: \rangle$ denote the average over all gauge field
configurations.

Fermionically disconnected components of these three-point functions
bring in additional contribution to the form factors, coming
from the quark-antiquark sea. 
At  zero-momentum transfer, such  contributions measure the charge
of the vacuum and therefore vanish. 
They also cancel-out at finite momentum-transfer,
if one assumes flavor SU(3) symmetry. 

\begin{figure}
\includegraphics[scale=0.7,clip=]{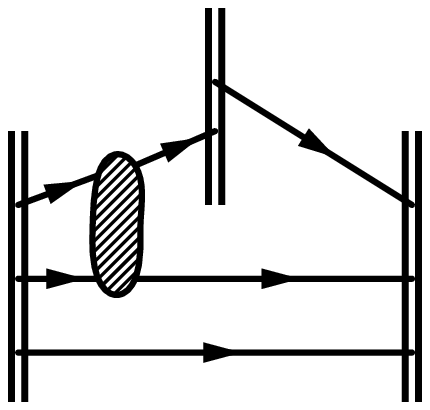}
\hspace*{1.5cm}
\includegraphics[scale=0.7,clip=]{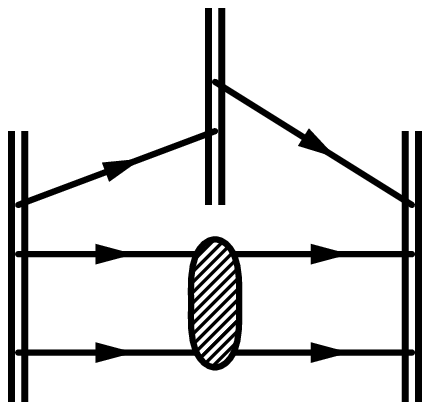}
\hspace*{1.5cm}
\includegraphics[scale=0.7,clip=]{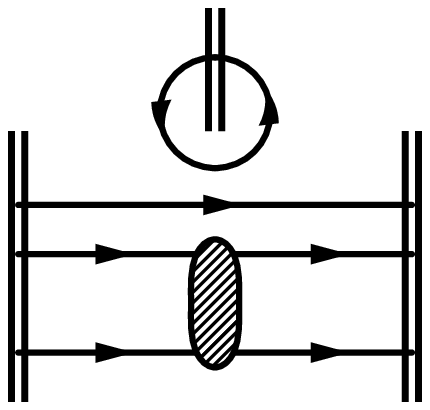} \\
(A) \hspace*{4.6cm} (B) \hspace*{4.6cm} (C) \\
\caption{Graphical representation of the typical contributions
to the W2W nucleon electro-magnetic three point
function. The double lined ``walls'' correspond to the
spatial Fourier integration.
The dashed ellipse denotes the four quark (zero-mode) instanton
interaction. The nucleon is excited at the left, struck by the
virtual photon in the middle and absorbed at the right. Two
contributions to the connected three-point function are shown.
Diagram~(A) probes the diquark content of the nucleon, whereas in
diagram~(B), the photon interacts with the remaining
quark. Diagram~(C) is disconnected, where the photon probes the sea quark
content of the nucleon.}
\label{singleI}
\end{figure}

So far, all expressions are completely general, as all the QCD dynamics resides
in the quantum average over the gauge configurations.
In the semi-classical limit, the non-perturbative
contribution to the correlation functions (\ref{G3E}) and (\ref{G3M})
arises from  single-instanton and from  many-instanton effects.
Typical single-instanton contributions
are represented in Fig. \ref{singleI}, where the instanton field mediates the
exchange of momentum between two partons.
Many-instanton effects are not only those
in which  which a parton exchanges its momentum with the other partons in the
nucleon by scattering on two or more pseudo-particles. 
In addition, there are also collective effects, 
which are associated with the breaking of chiral symmetry and
the  dynamical generation 
of a momentum-dependent quark effective mass \cite{dyakonov}.   
These many-body interactions
 are supposed to play an important role at low momenta.

In the next section, we shall calculate the 
contributions arising from the interaction of two massless
partons with a single-instanton, while 
many-instanton effects will be discussed in section \ref{MIGE}.

\section{ Single-Instanton contributions}
\label{siaFF}

In this section, we use the SIA to evaluate the single-instanton 
contribution to the correlators (\ref{G3E}), (\ref{G3M}) and (\ref{G2}).
The SIA is an effective theory of the instanton vacuum, in which
the degrees of freedom of the closest pseudo-particle are kept explicitly 
into account, while the contribution from all other pseudo-particles
 in the vacuum  is included into one effective parameter, $m^*\simeq$~85~MeV.
Such a parameter, which was rigorously defined and calculated in \cite{sia}
for different ensembles,
depends only on the two phenomenological parameters of 
the ILM, i.e. the instanton size $\bar{\rho}$ and density $\bar{n}$.

The main advantage of the SIA is that quark propagator in the 
single-instanton  back-ground  has a simple analytical form~\cite{brown78}.
It consists of a zero-mode part  and a 
non-zero mode part, $S^I(x,y)=S^I_{zm}(x,y)+S^I_{nzm}(x,y)$. 
The accuracy of the SIA was analyzed in detail in 
\cite{sia,3ptILM}. It was shown 
that the approach is reliable only if the relevant Green functions
receive maximal contribution from the zero-mode part of the 
propagator. In fact, the additional $\gamma_4$ matrix in
(\ref{G3E}), (\ref{G3M}) and~(\ref{G2}) has been inserted in order to 
meet such a requirement.

In this work, we choose to further simplify the calculation by
adopting the so-called "zero-mode approximation", in which the
non-zero mode part of the propagator is replaced by the free one,
$S^I(x,y) \simeq S^I_{zm}(x,y)+S_0(x,y)$. Such an approximation corresponds
to accounting for the 't~Hooft interaction, and neglecting other 
residual instanton-induced interactions, which are generally sub-leading.
Indeed, in ~\cite{3ptILM} it was shown that the zero-mode approximation is
very accurate in the case of the nucleon three- and two-point functions
which we are considering.

Finally, it is convenient to use the regular gauge and 
work directly in a time-momentum representation of the Green
functions.  To this end, one
expresses Eqs.~(\ref{G3wick}) and~(\ref{G2wick}) in terms of  ``wall-to-wall''
(W2W) propagators, defined as the spatial Fourier transforms of 
the point-to-point (P2P) quark propagators:
\be
\label{SW2W}
S(t',{\bf p'};t,{\bf p}) \equiv \int d^3{\bf x} \, d^3{\bf y} \,
e^{i \, {\bf p'} \cdot {\bf y} - i \, {\bf p} \cdot {\bf x}} \, S(y,x).
\ee
This is achieved by insertions of appropriate delta-functions at each
vertex.
The convenience of the time-momentum representation resides in the fact
that the W2W quark propagators in the single-instanton back-ground
have been calculated analytically~\cite{pionFF} and are smooth,
non-oscillatory exponential or Bessel functions.
The massless free W2W quark propagator is given by
\be
S_0(t',{\bf p'};t,{\bf p})=(2 \pi)^3 \, \delta^{(3)}({\bf p'}-{\bf p})
\, \frac{e^{-|{\bf p}|\,|t'-t|}}{2} \, u_\mu \gamma_\mu , \quad
\label{freeW2W}
\ee
where $u_4=-1$ and $u_l=-i \, p_l/|{\bf p}|$, for $l=1,2,3$.
The zero-mode W2W quark propagator in the regular gauge is given by:
\be
\label{zmW2W}
S^{I(A)}_{zm}(t',{\bf p'};t,{\bf p}) &=& \frac{2 \rho^2}{m^\star} \,
f(t',{\bf p'};t,{\bf p}) \, {\bf W}^{I(A)} , \\
f(t',{\bf p'};t,{\bf p}) &\equiv& e^{i \, ({\bf p'}-{\bf p}) \cdot
{\bf z}} \, K_0 \bigl( |{\bf p'}| \, \sqrt{(t'-z_4)^2+\rho^2} \bigr)
\,K_0 \bigl( |{\bf p}| \, \sqrt{(t-z_4)^2+\rho^2}
\bigr) , \\
{\bf W}^{I(A)} &\equiv& \gamma_\mu \, \gamma_\nu \,
\frac{1\pm\gamma_5}{2} \, \tau_\mu^\mp \,\tau_\nu^\pm ,
\ee
where $z_\mu=({\bf z},z_4)$ denotes the instanton position, $m^\star$ is the
effective parameter discussed above, and $\tau_\mu^\pm = ({\bf
\tau},\mp i)$ are color matrices.

The calculation of the correlation functions (\ref{G3E}), (\ref{G3M}) and
(\ref{G2})  is performed by substituting 
(\ref{freeW2W}) and (\ref{zmW2W}) in the 
traces arising from Wick contractions.
The quantum average is carried-out by integrating over
the instanton color orientation, position, and size.
The integral over the color orientation is trivial, while that over 
instanton position generates a delta-function which accounts for total 
momentum  conservation. As expected, the introduction of an instanton-induced
interaction  generates an extra loop-integral, over 
the momentum exchanged through the field of the instanton.
Despite the presence of loops, all diagrams are finite, as the instanton 
finite size
provides a natural ultra-violet cut-off.
The integral over the instanton size is weighted by a distribution 
function. In  this work we assume a simple delta-function distribution
$d(\rho)=\bar{n} \,\delta(\rho-\bar{\rho})$. 
Alternatively, one could use a fit of the instanton size distributions 
obtained  from lattice simulations 
(for a compilation of results see \cite{Negele}).
In a previous work we have verified that these two choices essentially
give the same result \cite{pionFF}.

The SIA is reliable only if the correlation functions 
are dominated by the contribution of the closest instanton. 
This condition is clearly not satisfied when the distance covered by the quarks
becomes much larger than the  typical distance between two neighbor instantons.
Previous studies \cite{sia,3ptILM} have shown that P2P
Green functions obtained analytically 
in the SIA quantitatively agree with those obtained numerically in the
full instanton back-ground, if the distance between the quark source and the
quark sink is smaller than $\sim~1$~fm for 
two-point functions and than $\sim~1.8$~fm, for three-point functions. 

On the other hand,  we do not expect the
SIA calculation of the W2W correlators
to be reliable for all values of the momentum $\bf{p}$,
even for small Euclidean times $t$.
In fact, if the momentum is small the spatial Fourier transform
(\ref{SW2W})
receives non-negligible contributions from  P2P propagators connecting
very distant points on the walls.
However,  for $|\bf{p}|$ larger than a GeV or so,  
only points at the distance smaller than roughly one inverse GeV
 from the time axis will contribute to the Fourier transform,
 and the SIA is applicable.

The physical reason why  at large $Q^2$ single-instanton effects dominate over
many-instanton contributions  is 
the following. In Minkowsky space, instantons correspond to
 quantum fluctuations
related to tunneling between degenerate classical vacua of QCD.
At large momentum transfer, one can imagine computing the
the form factor in an infinite-momentum frame, where the nucleon approaches
the speed of light. Following the same argument of Feynmann's parton model,
one concludes that in this frame the dynamics of the nucleon is frozen. 
As a result of such a time dilation, 
during the scattering process
quarks experience the consequences of - at most - a single tunneling event, 
i.e. of a single instanton.

In summary, the feasibility of SIA calculations
relies on the existence of a range of time and momentum, 
where the closest instanton contribution is dominant and the
 ground-state is isolated.
Previous studies \cite{sia,3ptILM,mymasses}
 have shown that, for the electro-magnetic three-point functions 
(\ref{G3E}) and (\ref{G3M}), 
this is achieved if one chooses $t$ to be
$0.8~\fm \lesssim t \lesssim~1~\fm$ and
restricts to the kinematic regime $|\bf{p}|\gtrsim~1-2~\textrm{GeV}$.

In order to compare SIA predictions against experiment
we shall first compute the nucleon coupling constant 
and mass from the two-point function.
Then, we shall use these values to extract the form factors from the 
three-point functions. All analytic results are collected in the appendix.

\subsection{Nucleon mass and coupling constant in the SIA}

In order to extract the nucleon
mass and coupling constant in the SIA, we have evaluated
the two-point function (\ref{G2})  for 
$t=0.9~\fm$ and $|{\bf p}|\gtrsim 1~\GeV$.
\begin{figure}
\includegraphics[scale=0.45,clip=]{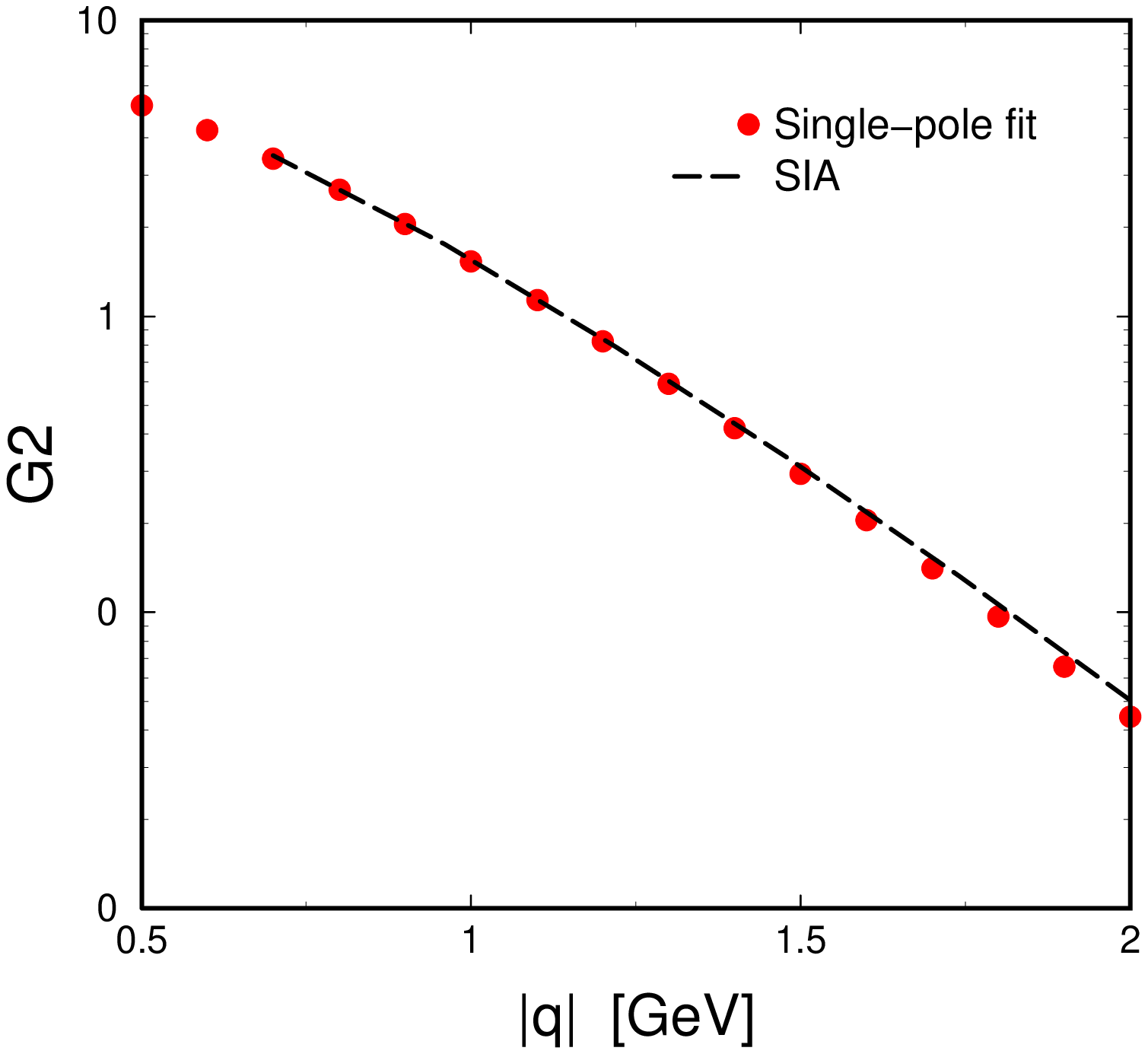}
\caption{(Color online) The nucleon two-point function, with $t$=0.9~fm (in units
of $10^6~GeV^9$).
The dashed line denotes the SIA prediction, the points 
represent the single-pole fit (\ref{spectralG2}), 
with $\Lambda_{\textrm{sc}}=0.030~\textrm{GeV}^3$ 
and $M=1.10$~GeV}
\label{twopoint}
\end{figure}

In Fig. \ref{twopoint} we show the SIA prediction for
${\bf G2}(t,{\bf q})$, and compare it 
with the single-pole fit from Eq. (\ref{spectralG2}), 
with $M=1.10$~GeV and $\Lambda_{\textrm{sc}}= 0.030~\textrm{GeV}^3$.
The agreement between SIA and the single-pole fit is very good, indicating 
that, for these values of time and momenta, the 
nucleon state has been isolated.

\subsection{Proton form factors in the SIA}

\begin{figure}
\includegraphics[scale=0.4,clip=]{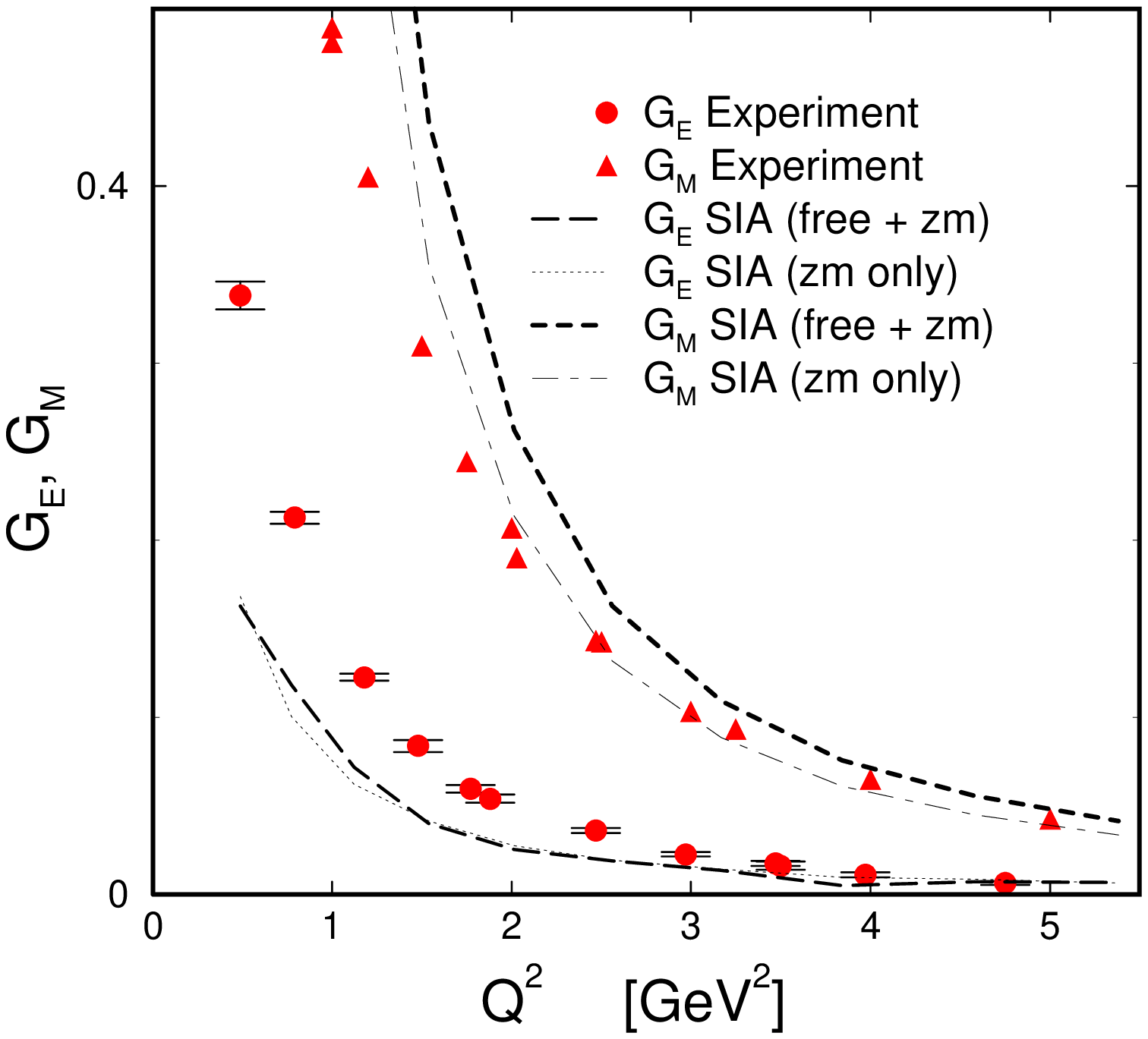}
\hspace*{1.5cm}
\includegraphics[scale=0.4,clip=]{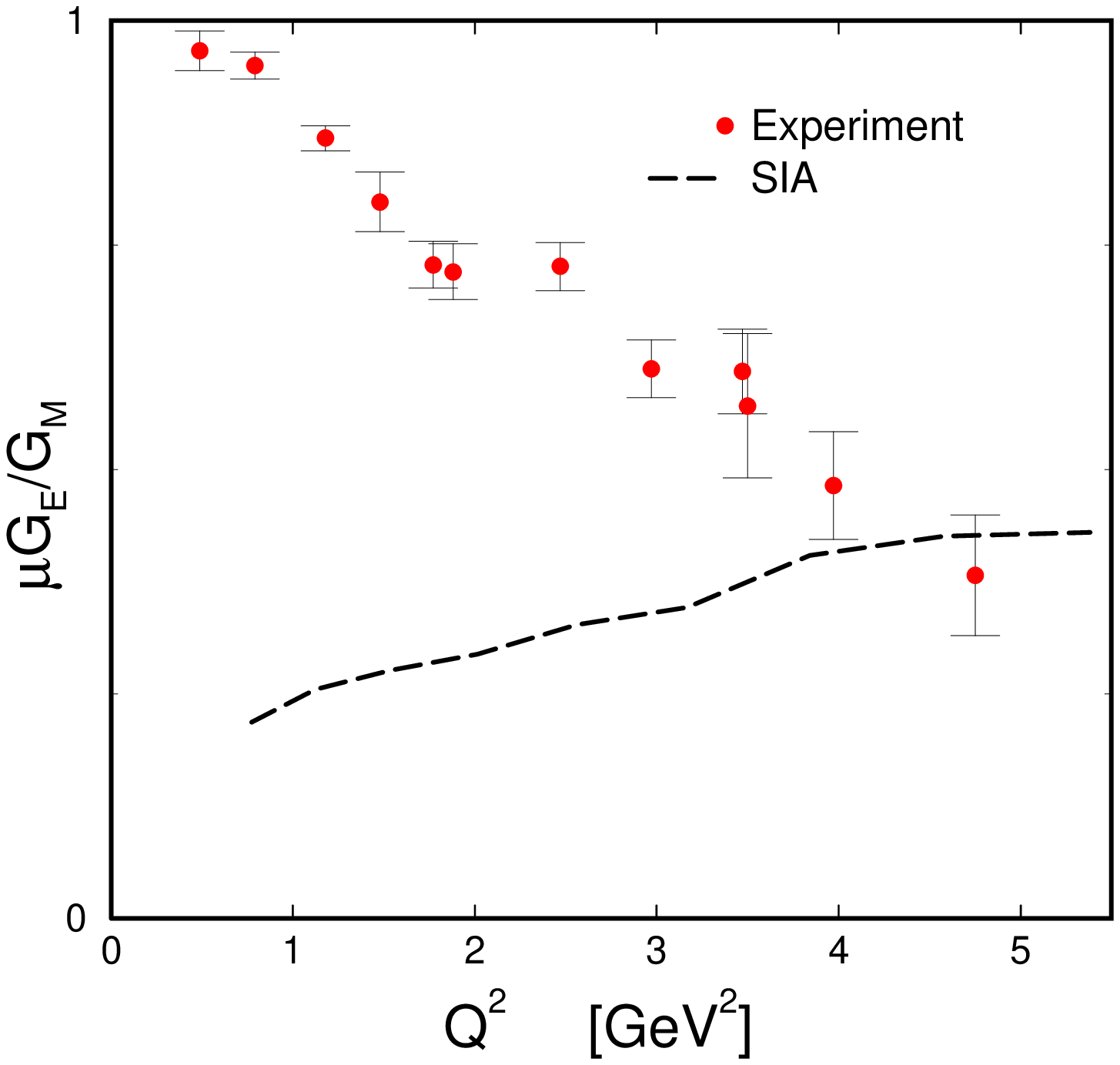}
{\large \qquad\qquad(A)\hspace*{7.5cm}(B)}
\caption{(Color online) $\mbox{}$\\
(A) SIA predictions for the proton Sachs form factors
compared to experimental data \cite{JLAB1,JLAB2,PGEGMexp}.
 The experimental points for 
electric form factor above $Q^2=0.5~\GeV^2$ are obtained from the
JLAB data for $\mu~G_E(Q^2)/G_M(Q^2)$, using a dipole fit for the
magnetic form factor. \\
(B) SIA prediction for the electric over magnetic form factors, compared to
recent JLAB data obtained by recoil polarization method~\cite{JLAB1,JLAB2}.}
\label{Sachs}
\end{figure}

After having extracted the nucleon mass and coupling, 
we are now in condition to discuss the single-instanton 
contribution to the proton form factors, which are obtained from 
the correlation functions (\ref{spectralG3E}),(\ref{spectralG3M}), 
(\ref{spectralG3F1}) and (\ref{spectralG3F2}), at\footnote{In 
\cite{nucleonGE} it was shown that, already for 
$t\gtrsim~0.7-1.0~\fm$,  
the relevant ratios of three- to two- point function  
are independent $t$.}  $t=0.9~\fm$. 
These theoretical predictions are affected by the errors 
generated by the numerical 
multi-dimentional loop integration and by the uncertainty 
on the best-fit values for $M$ and
$\Lambda_{\textrm{sc}}$.  The overall error
is estimated to be smaller than~$5\%$\footnote{
We note that the SIA prediction for $G_E$, reported in Fig. \ref{Sachs}A, 
does not exactly coincide with the results reported in \cite{nucleonGE}. 
Such a discrepancy is due to the fact that present
results are obtained with better numerical accuracy, which allowed us to 
determine more precisely the nucleon mass ($M=~1.10~\GeV~\pm~0.01~\GeV$ as
opposed to the early 
estimate $M=~1.17~\GeV~\pm~0.05$, used in \cite{nucleonGE}).}. 

The SIA results for the Sachs form factors of the proton are 
presented  in Fig.~\ref{Sachs}A and
compared to experimental data \cite{JLAB1,JLAB2,PGEGMexp}.
At relatively large momenta (~$Q^2\gtrsim~3~\GeV^2$~), where the
approach is supposed to work,  we observe  a good agreement between SIA
theoretical calculations and experiment.

In Fig. \ref{Sachs}B we show the single-instanton contribution to the
the ratio of magnetic and electric form factors.
Also in this case, we observe that theoretical calculations  converge 
toward the experimental data, in the large-momentum transfer regime.
However, we observe that at low-momentum transfer not only is the SIA
curve very far from experiment, but also its trend is opposite.

These results have several implications.
On the one hand, we find that single-instanton effects 
provide the right amount
of  non-perturbative short-distance dynamics needed 
to explain the observed Sachs form factors at
large momentum transfer.
On the other hand, we see that
the behavior of both the electric and the
magnetic form factors at
low- and intermediate-momentum transfer 
cannot be understood in terms of the interaction
of the partons with a {\it single} instanton.
In such a kinematic regime, form factors are expected to be 
very sensitive to many-instanton effects 
and possibly to other non-perturbative interactions.

\begin{figure}
\includegraphics[scale=0.40,clip=]{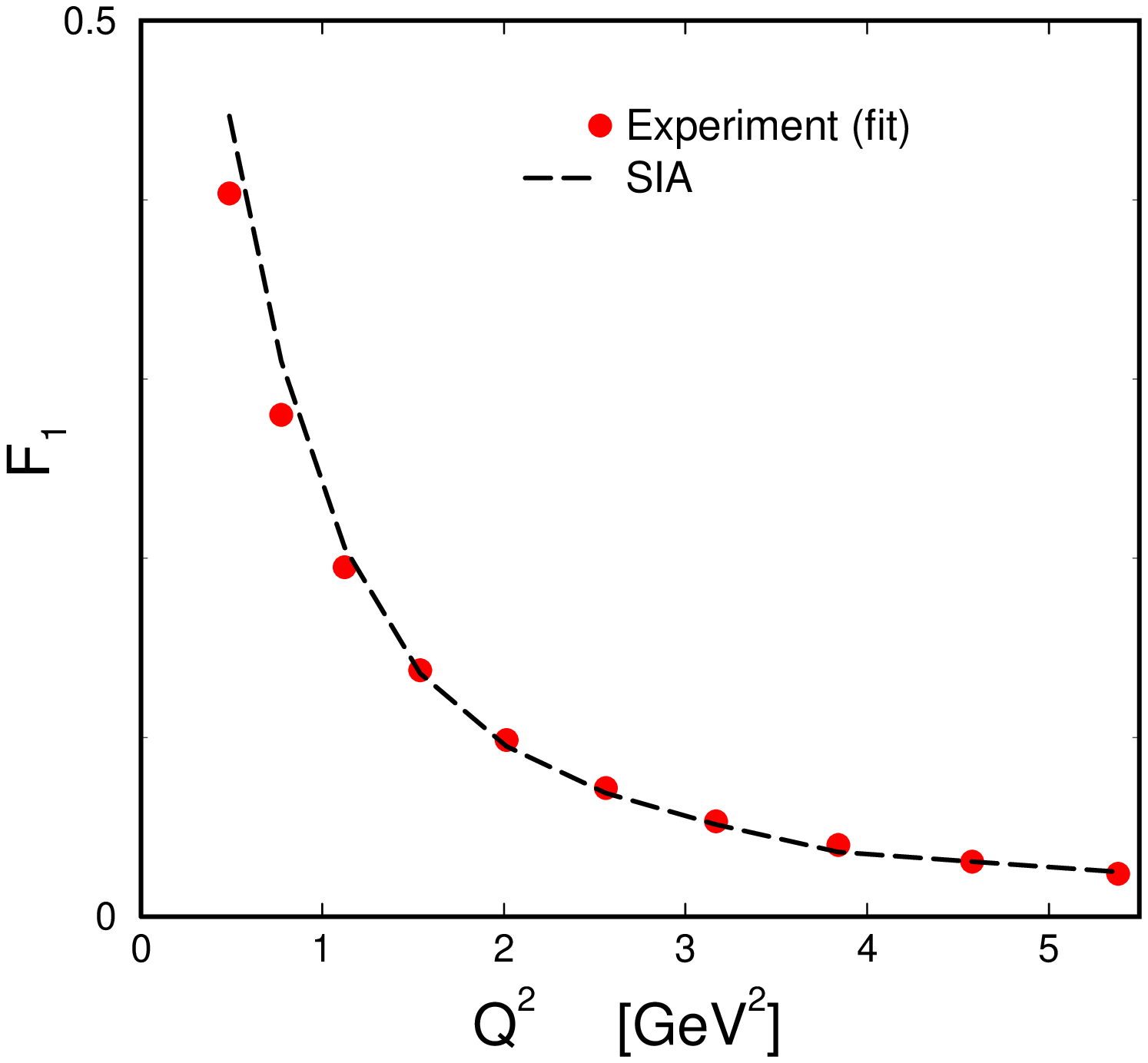}
\hspace*{1.5cm}
\includegraphics[scale=0.40,clip=]{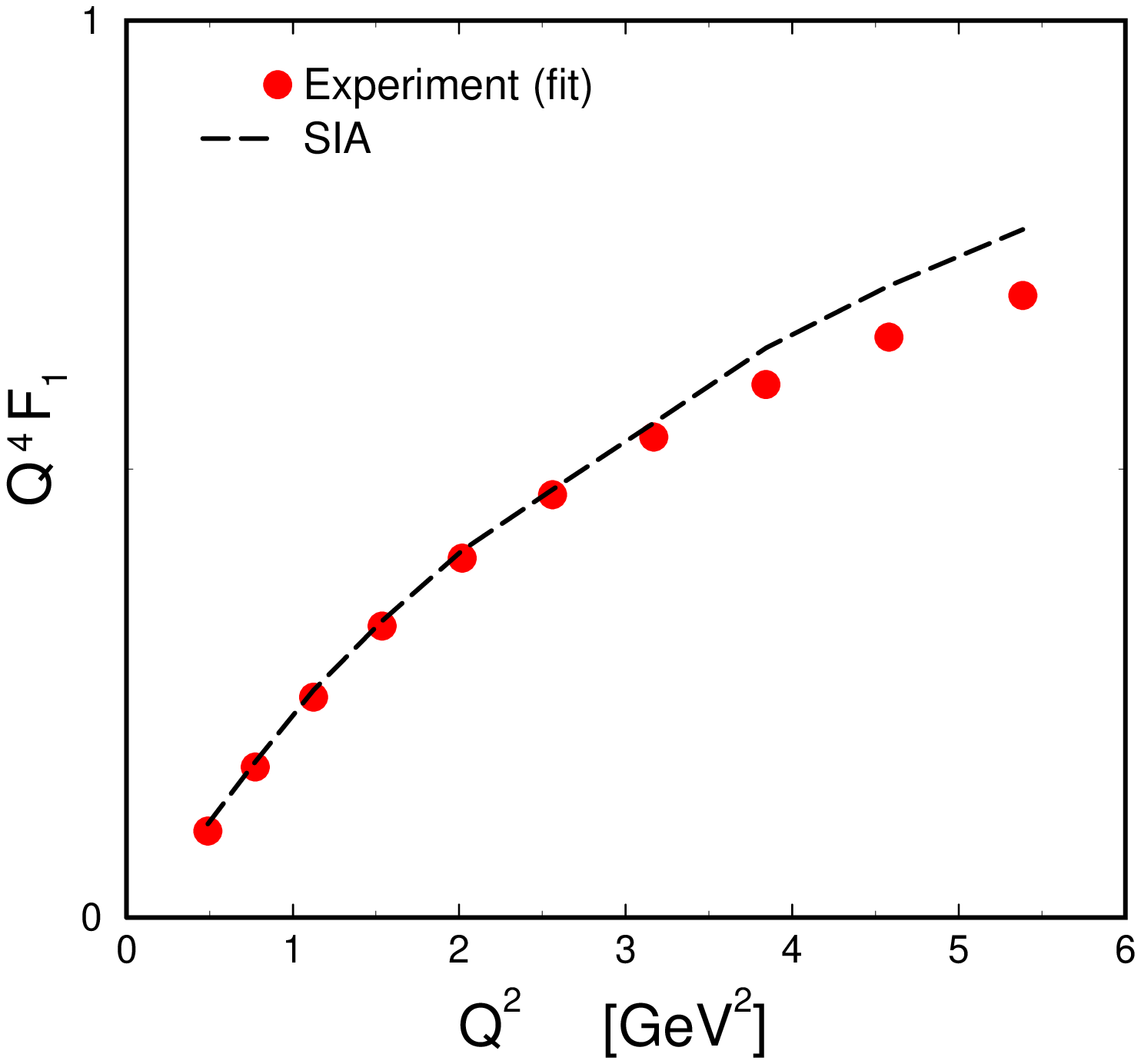}
{\large \qquad\qquad(A)\hspace*{7.5cm}(B)}
\caption{(Color online) $\mbox{}$\\
(A) Dirac form factor of the proton evaluated in the SIA and
compared to a phenomenological 
fit of the experimental data obtained as follows.
In (\ref{F1def}) the magnetic form factor is fitted with the traditional
dipole formula $\mu^{-1}~G_M^{fit}=G_{dip}(Q^2)=1/(1+Q^2/0.71)^2$. 
The electric form factor  is obtained from 
$G_E^{fit}(Q^2)=G_{dip}(Q^2)\times(1.-0.13\,(Q^2-0.04))$, 
where the second factor parametrizes the JLAB data for
$\mu\,G_M/G_E$. \\
(B) $Q^4$ times the Dirac form factor in the SIA and compared to a 
phenomenological fit of the experimental data. Perturbative
QCD counting rules predict $Q^4\,F_1(Q^2)\sim$~const.}
\label{Diracp}
\end{figure}

\begin{figure}
\includegraphics[scale=0.4,clip=]{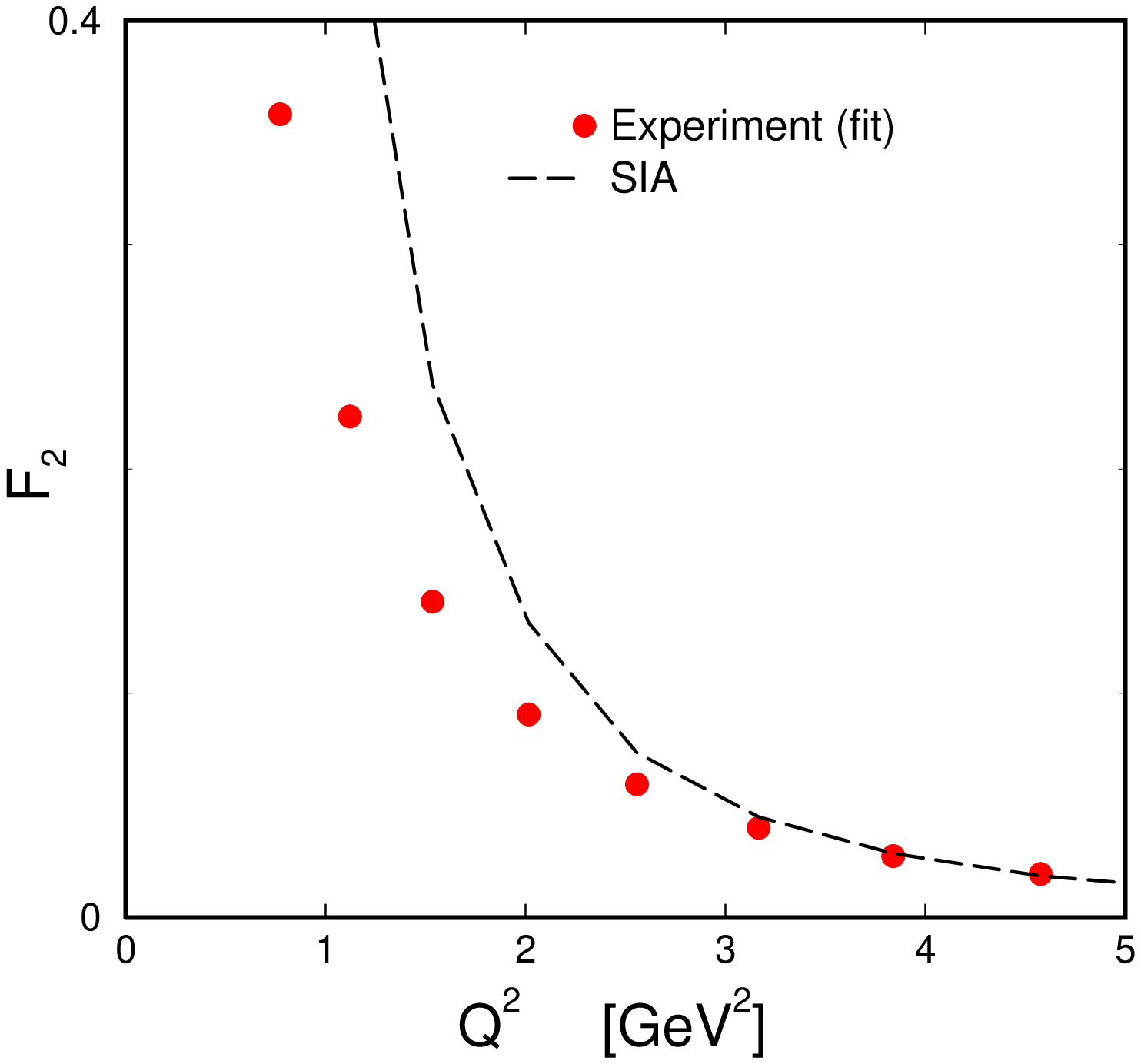}
\hspace*{1.5cm}
\includegraphics[scale=0.4,clip=]{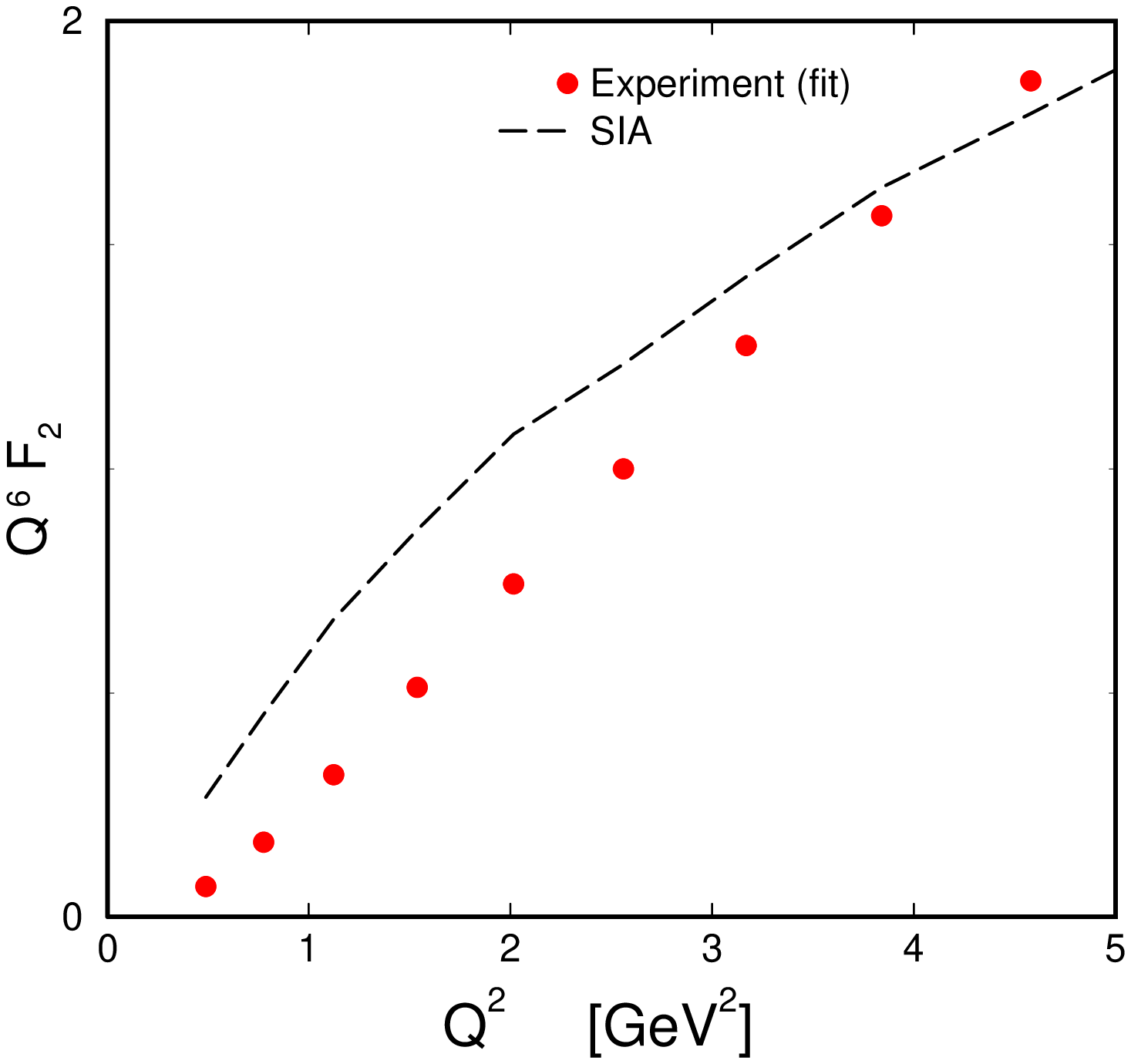}
{\large \qquad\qquad(A)\hspace*{7.5cm}(B)}
\caption{(Color online) $\mbox{}$\\
(A) Pauli form factor of the proton evaluated in the SIA and
compared to a phenomenological 
fit of the experimental data obtained as follows.
In (\ref{F1def}) the magnetic form factor is fitted with the traditional
dipole formula $\mu^{-1}\,G_M^{fit}=G_{dip}(Q^2)=1/(1+Q^2/0.71)^2$. 
The electric form factor  is obtained from 
$G_E^{fit}(Q^2)=G_{dip}(Q^2)\times(1.-0.13\,(Q^2-0.04))$, 
where the second factor parametrizes the JLAB data for
$\mu\,G_M/G_E$. \\
(B) $Q^6$ times the Pauli form factor of the proton
 in the SIA and compared to a 
phenomenological fit of the experimental data. Perturbative 
QCD at lowest-twist predicts
$Q^6\,F_2(Q^2)\sim$~const, modulo logarithmic corrections.}
\label{Paulip}
\end{figure}

Let us now discuss the  SIA results for the Dirac and Pauli form factors,
which are reported in Fig.s (\ref{Diracp}) and (\ref{Paulip})
and compared with the fit of the experimental results.
We observe that single-instanton effects are sufficient to 
explain with impressive accuracy the Dirac form factor, from low- to high- 
$Q^2$. Notice that, at the largest momentum available $Q^2\simeq~5.6~\GeV^2$,
the slope of the function $Q^4~F_1(Q^2)$ is still  larger than zero.
On the other hand, we recall that in pQCD this combination should
 be a constant,   modulo logarithmic corrections.
Hence, we conclude that single-instanton effects provide the right amount 
of dynamics required to explain the deviation from the perturbative behavior
of the Dirac form factor.

The SIA prediction for the proton Pauli form factor is reported in 
Fig.~\ref{Paulip} and compared to a fit of the experimental data.
In this case, the performance of the SIA at low-momentum transfer
is worse than in the case of the Dirac form factor.

It is natural to ask why the same approach performs differently in the two 
cases.
We recall that the SIA is an effective theory of the ILM which
can be used to account for instanton effects 
only in the limit of large momentum transfer.
Therefore, the fact that the SIA prediction deviates from the data at
small momentum tranfer  does not necessarily 
imply that the instanton model is in disagreement with experiment. 
In order to check the ILM against 
low-energy experimental data one necessarily needs to perform a 
many-instanton calculation.

With this in mind, let us 
compare the definitions of the Dirac and Pauli form factors, in terms of 
three-point correlation functions, Eq.s (\ref{spectralG3F1}) and 
(\ref{spectralG3F2}).
We observe that the $F_2(Q^2)$ is obtained from a {\it difference} of
correlation functions of comparable magnitude
(recall that ${\bf G3}^{p(n)}_{F_2}$ is negative
definite), while $F_1(Q^2)$ is related to the {\it sum} 
of the same quantities. 
Notice also that in the combination leading to $F_2(Q^2)$
the contribution of the magnetic correlator is weighted by
the inverse of $Q^2$, (through the factor $1/\tau$) 
which enhances the low-momentum modes, for which the SIA becomes inaccurate. 
From this observations it follows that
 the systematic error caused by the use of the 
SIA in the intermediate- and low- momentum
regime is larger in the case of the Pauli form factor than in the case of 
the Dirac form factor.

In the present calculations, 
all perturbative fluctuations have been neglected. It is therefore
important to have at least an estimate of the magnitude of these 
contributions. 
To this end,  in Fig. \ref{Sachs}A  we compare
the complete SIA results with the predictions obtained by retaining only the
zero-mode part of the propagator. The difference between these two curves
comes from {\it free} diagrams. 
By definition of perturbation theory, the contribution from free diagrams
has to be larger than the perturbative corrections to them.
So by comparing free versus zero-mode contributions, we can estimate the
importance of perturbative fluctuations, relative to the non-perturbative
effects we have accounted for. 
In all cases considered, we found that the 
instanton-induced contributions represent the dominant dynamical effect. 

In summary, we have observed that SIA is able to reproduce the Sachs form
factor in the regime where it is applicable, i.e. 
at large momentum transfer.
On the other hand, the approach misses important
dynamics in the low- and intermediate-momentum transfer, where many-instanton
effects have to be included.
Interestingly, we have observed that such many-body contributions
are not important in the Dirac form factor, which 
is extremely well reproduced in the SIA, from rather small to
large $Q^2$.

\subsection{Neutron Form Factors in the SIA}

\begin{figure}
\includegraphics[scale=0.4,clip=]{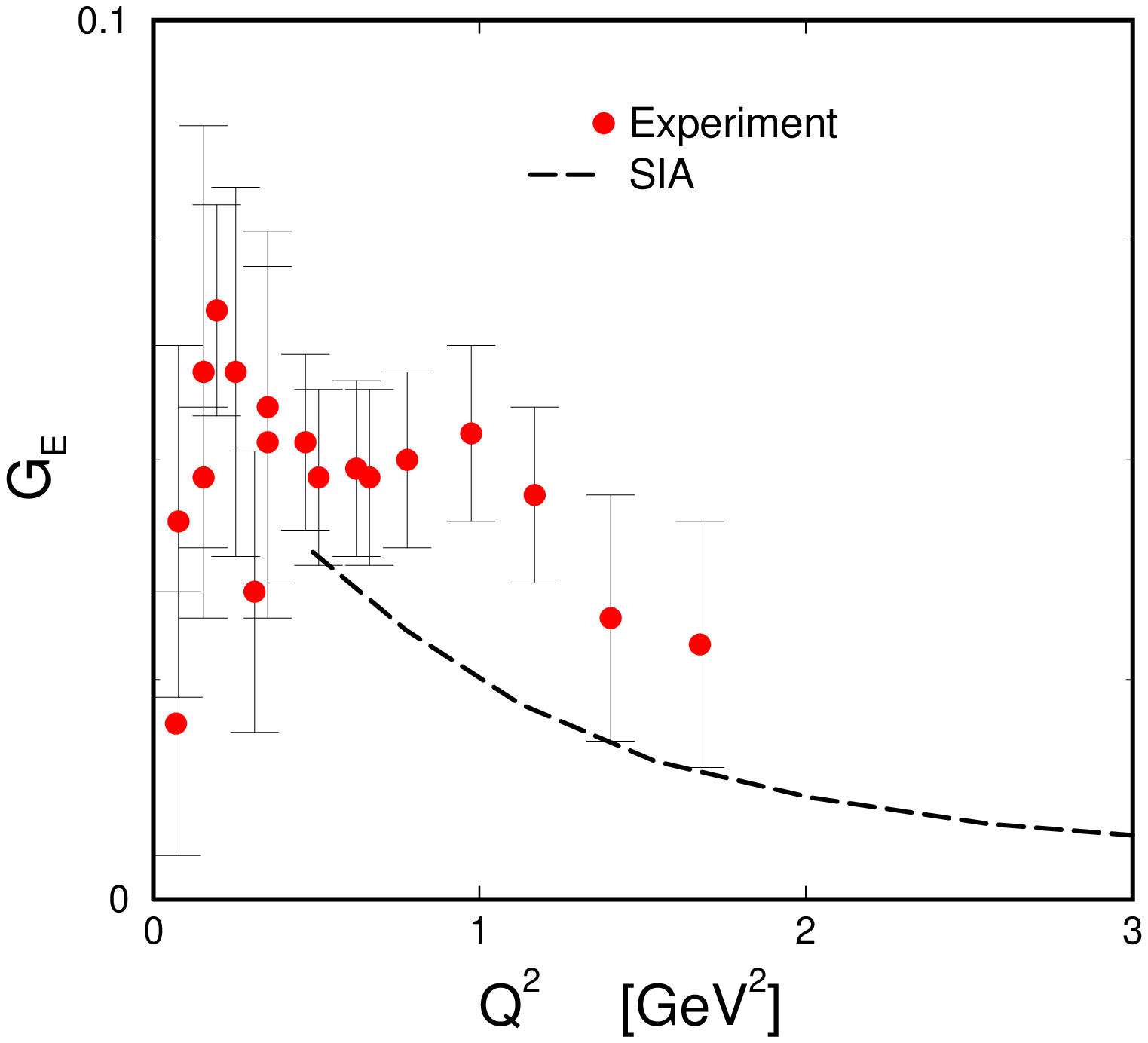}
\hspace*{1.5cm}
\includegraphics[scale=0.4,clip=]{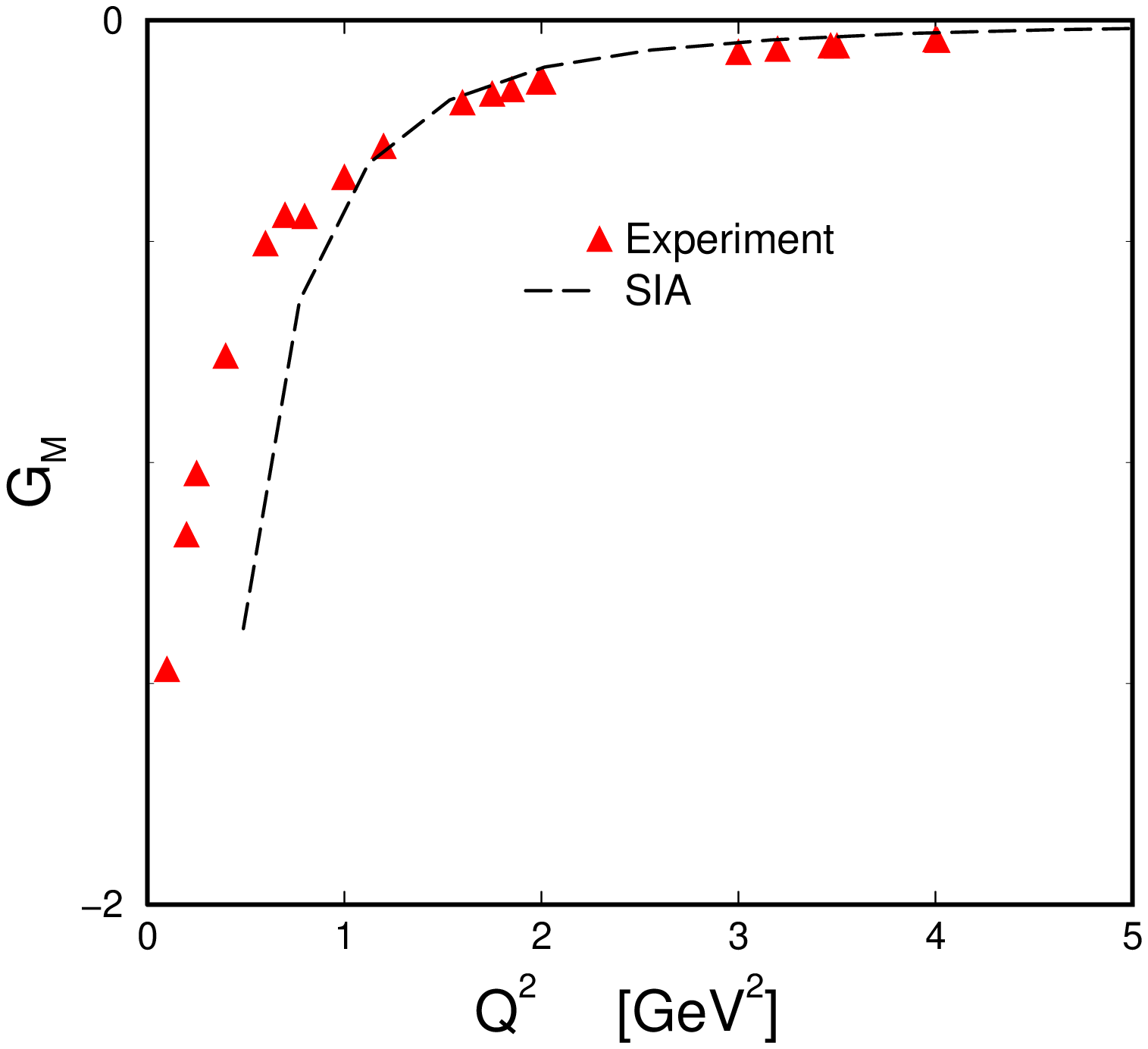}
{\large \qquad\qquad(A)\hspace*{7.5cm}(B)}
\caption{(Color online) $\mbox{}$\\
(A) Electric form factor of the neutron evaluated in the SIA and
compared to experimental data~\cite{NGEexp}.\\
(B) Magnetic form factor of the neutron evaluated in the SIA and
compared to  experimental data~\cite{NGEexp}.}
\label{Sachsn}
\end{figure}

The result of the SIA calculations of the neutron electro-magnetic 
form factors  are presented in Fig. \ref{Sachsn} 
and compared with the experimental data. 

As in the case of the proton, we observe that single-instanton effects
can explain the data on magnetic form factor in the large 
momentum transfer regime.
On the other hand, the electric form factor is known only at small 
momentum transfer, where the SIA is not reliable. In this case, the SIA 
undershoots the experimental data by a factor two or so.
Clearly, in order to 
test the validity of the ILM with such a form factor, we need to include
many-instanton effects.

\begin{figure}
\includegraphics[scale=0.4,clip=]{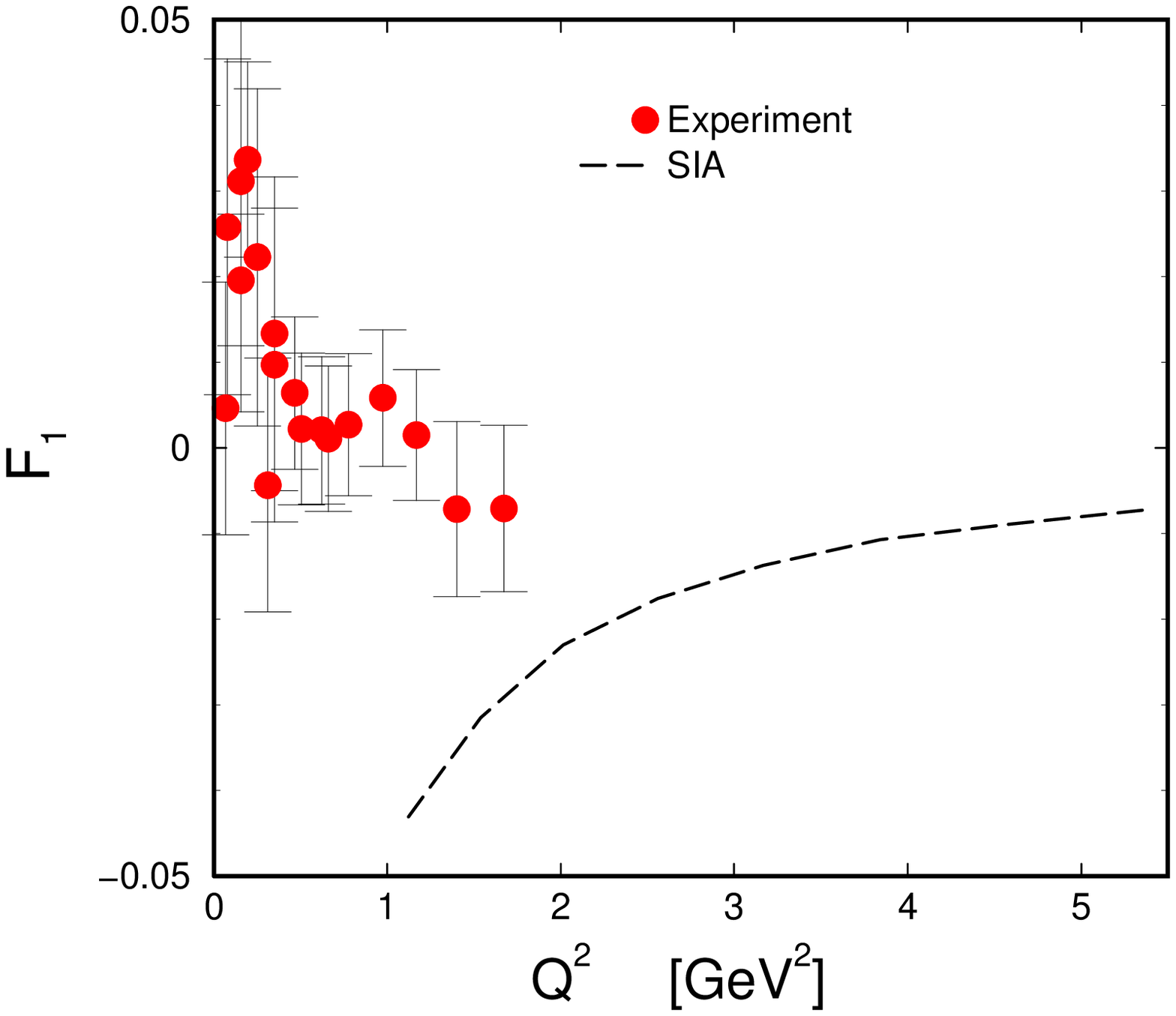}
\hspace*{1.5cm}
\includegraphics[scale=0.4,clip=]{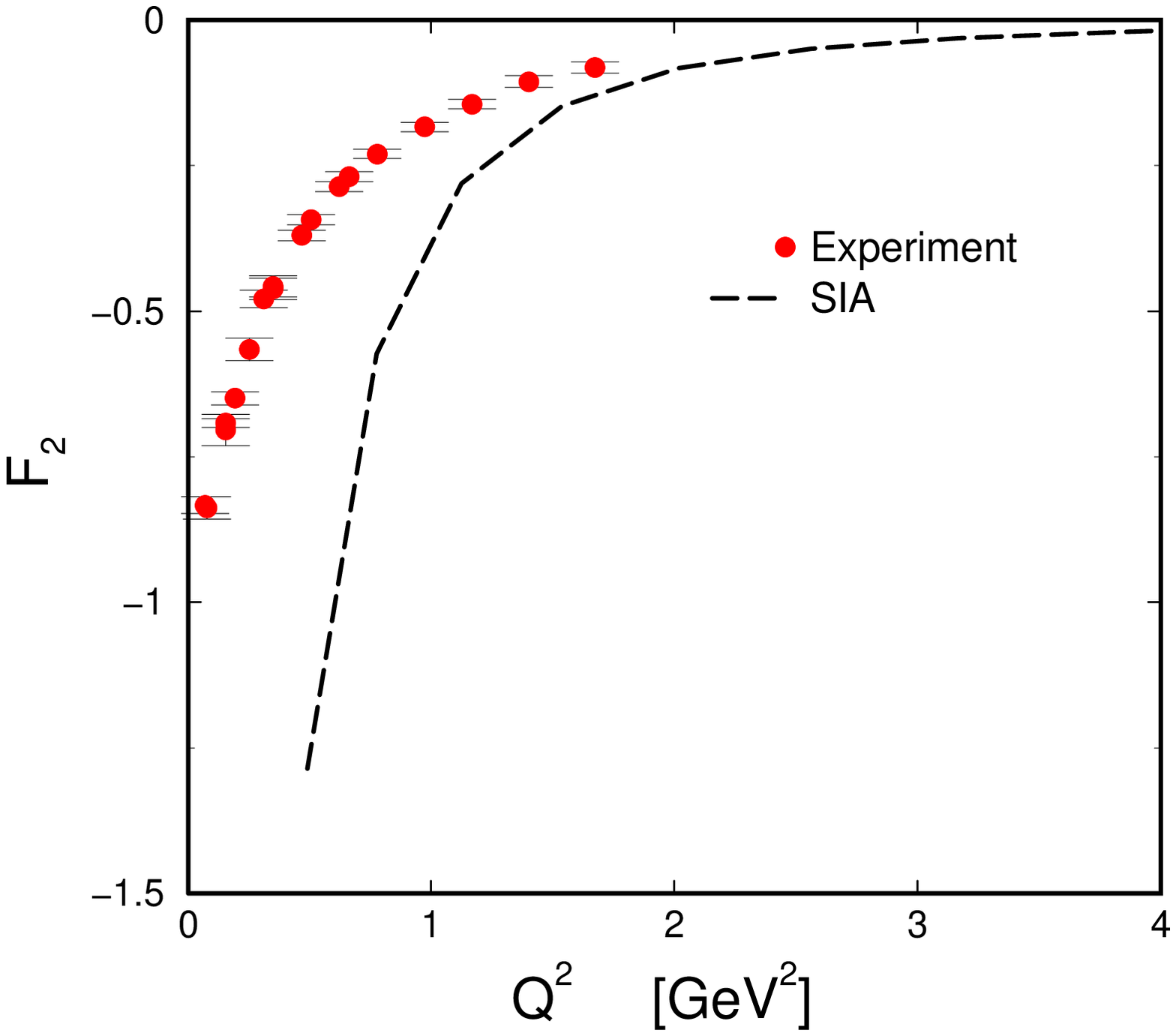}
{\large \qquad\qquad(A)\hspace*{7.5cm}(B)}
\caption{(Color online) $\mbox{}$\\
(A) Dirac form factor of the neutron evaluated in the SIA and
compared to experimental data.
The experimental curve has been obtained by assuming 
that the magnetic form factors 
follows a dipole formula and taking the electric form factor form experiment\\
(B) Pauli form factor of the neutron evaluated in the SIA and
compared to experimental data.
The experimental curve has been obtained by assuming 
that the magnetic form factors 
follows a dipole formula and taking the electric form factor form experiment.}
\label{paulidiracn}
\end{figure}

The SIA predictions for the neutron Pauli and Dirac form factors, which
are also known only at small momentum transfer,  are presented
for completeness in Fig. \ref{paulidiracn} and compared against 
experimental data.
In these cases, we observe that the agreement between SIA and these low-energy
data is indeed quite poor. 

In general, we have found that
single-instanton effects alone are not sufficient to explain 
the available low-energy  information on the 
form factors of the neutron.

\section{Many-Instanton Contributions}
\label{MIGE}
In the previous section we have analyzed the single-instanton contribution
to the form factors of the nucleon. In general, we observed a good agreement 
with  experimental data, in the large momentum transfer regime.
On the other hand, we have verified that at low-momentum transfer 
the single-instanton effects are sub-leading, as expected.
Thus, in order to address the question whether also the low-energy data can be 
explained by the 't~Hooft interaction, we need to account for 
many-instanton degrees of freedom explicitly.
To do so, we face the problem of computing 
the relevant correlation functions  in the full instanton liquid vacuum, 
i.e. to \emph{all orders}  in the 't~Hooft interaction. 

Such ILM calculations can be 
performed by exploiting the analogy between the Euclidean generating 
functional and the partition function of a statistical 
ensemble~\cite{shuryakrev}, in close 
analogy with what is usually done in lattice simulations.
After the integral over the fermionic degrees of freedom is carried out
explicitly, one computes
expectation values of the resulting Wick contractions (\ref{G3wick}) by
performing a Montecarlo average over the configurations of an ensemble of
instantons and anti-instantons. In the Random Instanton Liquid (RILM), 
the density and size of the pseudo-particles is kept fixed,
while their position in a periodic box and their 
color orientation is generated according to a random
distribution.

In this framework, P2P correlators can be evaluated 
accurately in a few hours on a regular work-station. 
Unfortunately, the W2W correlators which are needed in order to
extract the form factors are much harder to compute numerically.
Indeed, many simplifications which make the SIA approach particularly 
convenient do not occur in a multi-instanton back-ground. 
For example, since at the one-instanton level the W2W  quark propagator 
in the instanton back-ground is known in a closed form, 
 one can carry out 
calculations analytically, working directly in a time-momentum representation. 
On the other hand, in a multi-instanton back-ground the quark propagator is
obtained by inverting numerically the Dirac operator, and this 
operation is done in coordinate representation. 
Hence, one is left to computing numerically the six-dimensional integration
in  Eq.s  (\ref{G3E}) and (\ref{G3M}). 
Furthermore, such an integration is complicated by the nasty oscillatory
behavior of the integrand, introduced by the phases of the Fourier transform. 

As a result of these facts, while P2P correlators
can be evaluated on an ordinary single-processor computer,
W2W correlators typically call 
for a multi-processor computation.
But even on a very powerful parallel machine, an accurate evaluation of the 
form factors at large momentum transfer is still very hard to achieve,
because in such a kinematic regime the integrand is oscillating very fast. 
In this section, we propose a strategy to overcome these problems.

We begin by analyzing the many-instanton contribution
to the electric form factors, for which 
an important simplification occurs, as we shall see below. 
As a first step,  we rewrite (\ref{G3E}) as:
\be
{\bf G3}^{p(n)}_{E}(t,{\bf q},{\bf P}) =
\int d^3{\bf x} \, d^3{\bf y} \, e^{-i \, {\bf q} \cdot {\bf y} +i \,
{\bf P} \cdot{\bf x}} \, \langle 0 | \, \text{Tr} \,
[\,\eta_{\text{sc}}^{p(n)}(2\,t,{\bf 0}) \, J_4^{em}(t,{\bf y}) \,
\bar{\eta}_{\text{sc}}^{p(n)}(0,{\bf x}) \, \gamma_4 \,] | 0 \rangle.
\nonumber\\
\label{G3ERF}
\ee
Note that charge conservation implies the identity:
\be
{\bf G3}^{p(n)}_E(t,{\bf 0},{\bf P})={\bf G2}(2 t, {\bf P}) 
\label{Qcons}
\ee
which can be useful to test the accuracy of the numerical integration. 
We can now eliminate one of the complex phases by setting ${\bf P}=0$, 
which corresponds to going to the nucleon's rest frame.
At large Euclidean times, the resulting
 Green's function has the following spectral  representation:
\be
\label{spectralG3ERF}
{\bf G3}^{p(n)}_E(t,{\bf q};{\bf 0}) 
 &\rightarrow& 
\Lambda^2_{\text{sc}}\,
\frac{\omega_{\bf{q}}+M}{\omega_{{\bf q}}} \,
e^{-\,\omega_{{\bf q}} \, t} \,e^{-M \, t}
G^{p(n)}_{\text{E}}(Q^2).
\ee
Now we observe that two of the three integrals in $d^3{\bf y}$ can be
performed analytically, exploiting the fact that the above Green's function
 is invariant under spatial rotations\footnote{Notice that only the electric 
three-point  function and the two-point function display such a symmetry 
property. Hence the method presented in this section cannot be applied
 to compute the magnetic form factor.}. We obtain:
\be
{\bf G3}^{p(n)}_{E}(t,{\bf q},{\bf 0}) =
\frac{4\,\pi}{|{\bf q}|}\,\int\, d|{\bf y}|\, |{\bf y}|
\sin(|{\bf q}|\,|{\bf y}|)\,{\bf \Gamma3}_E^{p(n)}(t, |{\bf y}|, {\bf 0}),
\label{G3Esymm}
\ee
where we have introduced the ``charge distribution Green's function'':
\be
{\bf \Gamma3}_E^{p(n)}(t, {\bf y},{\bf P})=
\int \frac{d^3{\bf k}}{(2\,\pi)^3}\,{\bf G3}_E^{p(n)}(t, {\bf k},{\bf P})
\,e^{i\,{\bf k}
\cdot {\bf y}}.
\ee
${\bf\Gamma3}^{p(n)}(t,{\bf y},{\bf 0})$ represents 
the probability amplitude for one of the three quarks which  
were created at an initial time in a state with 
with quantum numbers of the proton (neutron) and vanishing total momentum,
to absorb  a photon at a distance ${\bf y}$ from the origin of 
the center of mass frame, at a later time $t$.
When the Euclidean time becomes large such a Green function 
encodes the information about the charge distribution of the nucleon.

\begin{figure}
\includegraphics[scale=0.4,clip=]{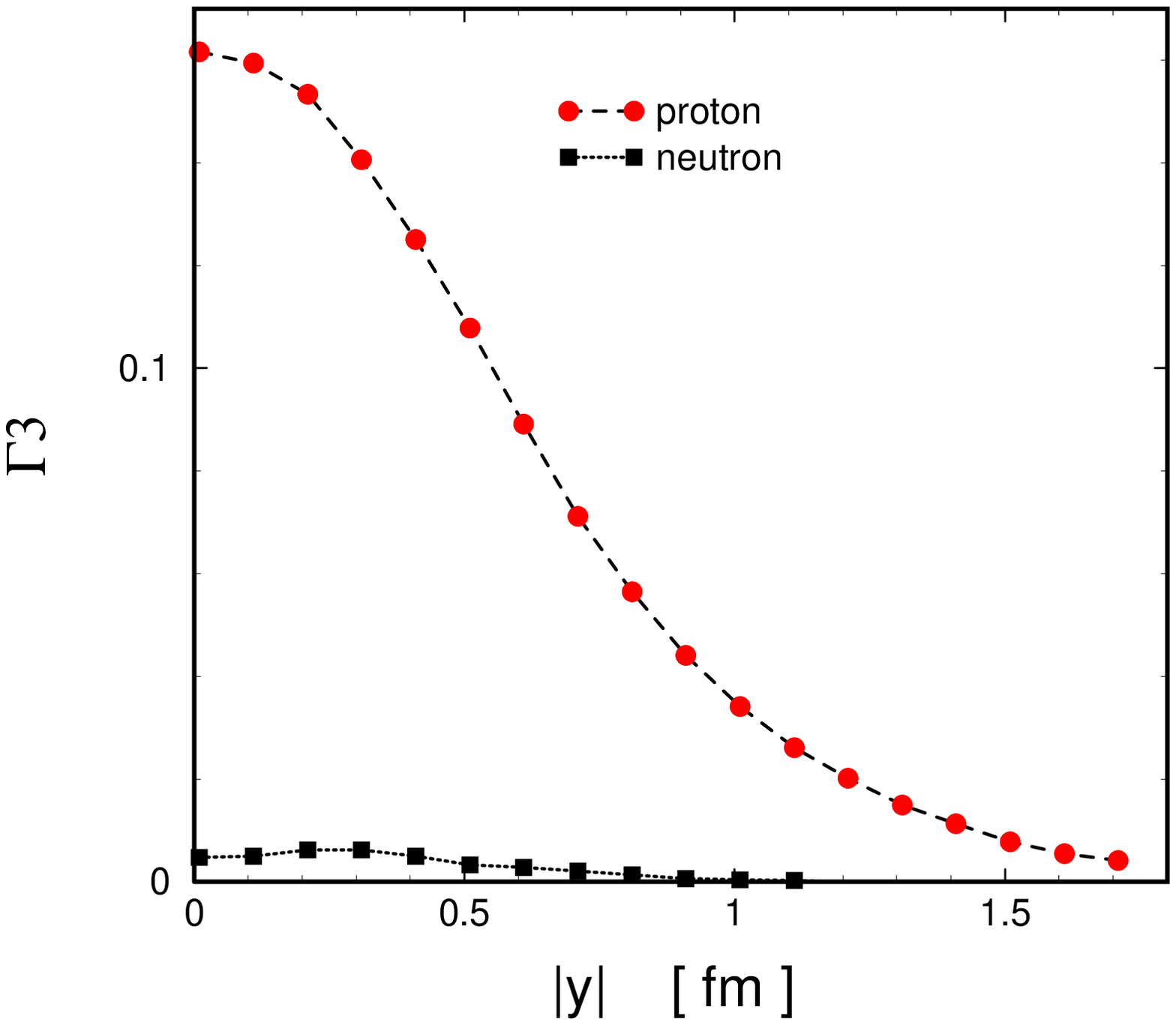}
\hspace*{1.5cm}
\includegraphics[scale=0.4,clip=]{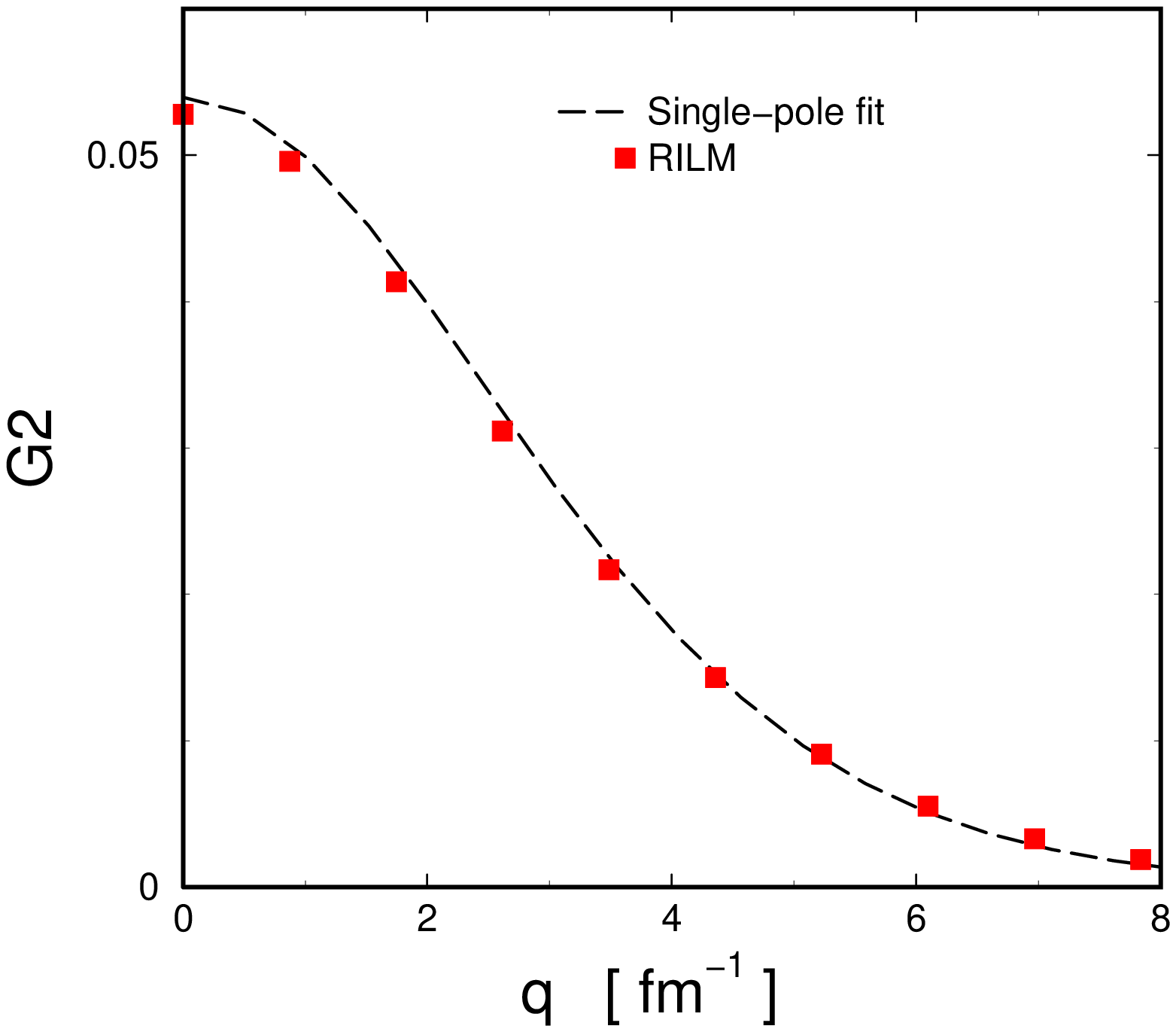}
{\large \qquad\qquad(A)\hspace*{7.5cm}(B)}
\caption{(Color online) $\mbox{}$\\
(A): The charge distribution Green's function
${\bf \Gamma 3}(t,|{\bf y}|,{\bf 0})$ for proton (circles) and 
nucleon (squares), evaluated
in the RILM for $t=0.9~\fm$.\\
(B): Two-point function of the nucleon in the RILM (points),
 compared to a single-pole fit (dashed line),
 ${\bf G2}_{fit}(t,|{\bf q}|)=2\,\Lambda^2\,e^{-t\,\sqrt{{\bf q}^2+M^2}}$ 
with $t=0.9~\fm$.}
\label{Gamma}
\end{figure}

Calculating numerically 
${\bf \Gamma3}^{p(n)}(t,|{\bf y}|,{\bf 0})$ for several values of
$|{\bf y}|$ is not computationally very challenging, 
because it requires only a three-dimensional
integration over the spatial position of the source and involves
no oscillating phase. This problem can be handled 
with traditional adaptive 
Montecarlo methods, and takes a few days of computation
on a  regular single-processor machine.
Then, for the final integration in $d\,|{\bf y}|$, we can make use of the
one-dimensional integration routines which are optimized for 
fast-oscillating functions.

We have evaluated the function ${\bf \Gamma3}(t,{\bf y},{\bf 0})$ 
in the RILM  by averaging over
configurations of 252 pseudo-particles of size $\rho=0.33~\fm$,
in a periodic box\footnote{As usual, in a finite box
all momenta are quantized according to 
$p_i=\frac{\pi}{L_i}\,n$, with $ n=0,\pm 1,\pm 2,...$.} of volume 
$(3.6^3\times 5.4)~\fm^4$.
Like in lattice simulations, we have used a rather large current quark mass 
(70~MeV), in order to avoid finite-volume artifacts.
The results for ${\bf \Gamma3}(t,|{\bf y}|,{\bf 0})$ for different values
of $|{\bf y}|$  are plotted in Fig (\ref{Gamma})~A.
The final one-dimensional integration in (\ref{G3Esymm}) has be handled
with a Gauss quadrature routine,
 combined with a polynomial interpolation of the integrand.

In order to extract the form factor, we have adopted the ratio of three- and
two-point functions  similar to the one suggested in \cite{draper}:
\be
G_E(Q^2) \lim_{t\to \infty}\frac{2~\omega_q}{M+ \omega_q}
\frac{ {\bf G3}_E(t,{\bf q},{\bf 0})}
{{\bf G2}(2t, {\bf q})}\,
\frac{{\bf G2}(t,{\bf q})}{{\bf G2}(t, {\bf 0})}.
\ee
\subsection{Nucleon Mass in the ILM}
As in the previous SIA calculation, before extracting the form factor
 we need to verify  that, at the
Euclidean time we work at ( $t=0.9~\fm$ ), the contribution of the
nucleon pole to the two-point function has been isolated.
To this end, in Fig. \ref{Gamma}~B we compare our numerical results in 
the RILM with a single-particle fit from (\ref{spectralG2}).
The mass extracted from the fit is $M=1.15~\GeV$, in good agreement with
previous estimates in the RILM \cite{corrbaryons,mymasses}.

\subsection{Proton Form Factors in the ILM}

\begin{figure}
\includegraphics[scale=0.45,clip=]{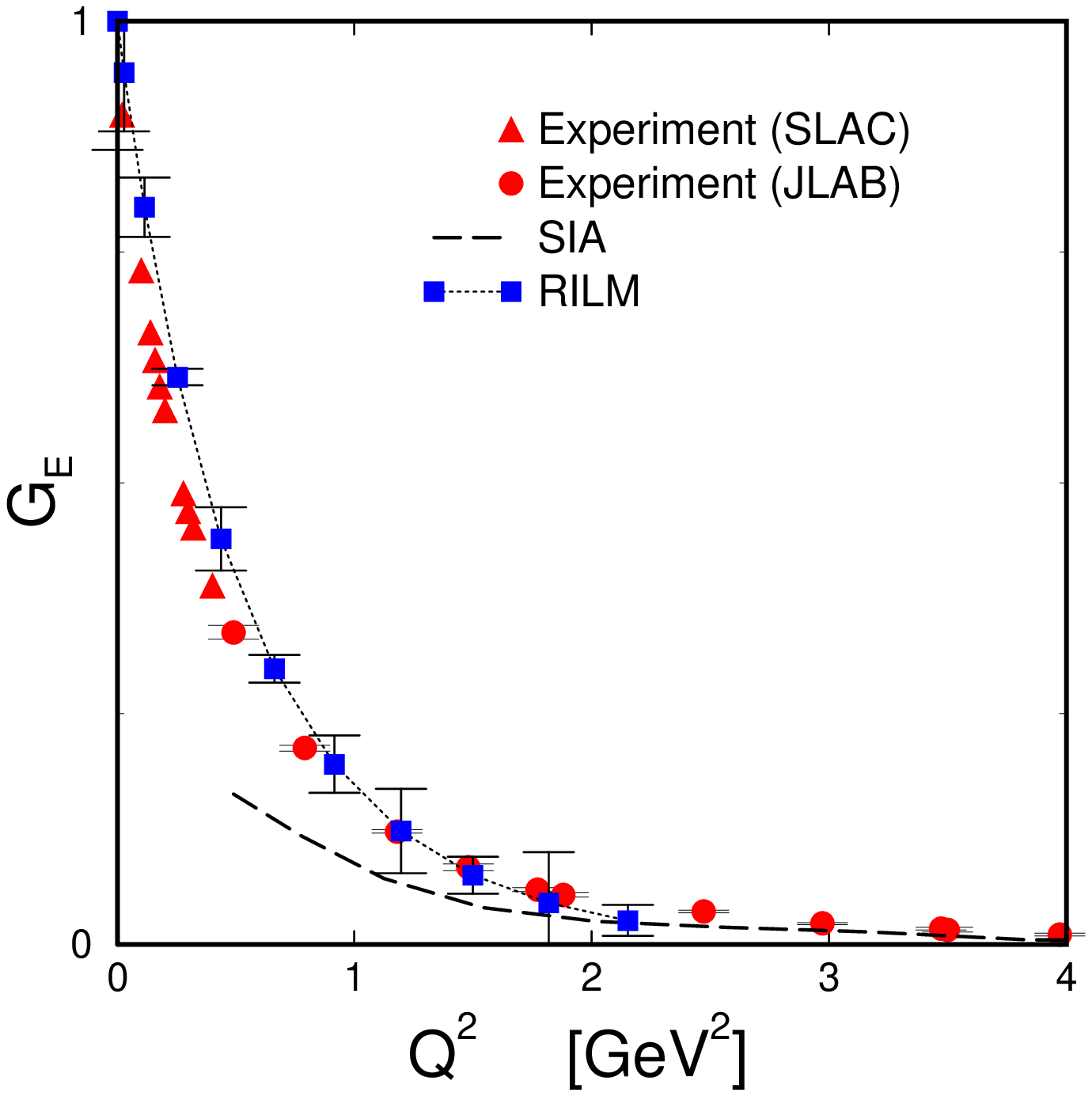}
\caption{(Color online) $\mbox{}$\\
(A): Electric form factor of the proton in the ILM and from experiment.
Triangles are low-energy SLAC data, which follow a dipole fit.
Circles are experimental data obtained from the
recent JLAB result for $G_E/G_M$, assuming a dipole fit for the 
magnetic form factor.
Squares are result of many-instanton simulations in the RILM, and the
dashed line is the SIA curve.}
\label{GEfig}
\end{figure}

The result of our calculation of the proton electric form factor in the RILM
is presented in Fig. \ref{GEfig}, where it is compared with 
experimental data and with the SIA curve.
We observe a very good agreement between theory and experiment. 
In particular, the inclusion of many-instanton effects allows to explain
the experimental data in the low-momentum regime, while 
at large momentum transfer  the RILM gives
results completely consistent with the simple single-instanton calculation.
Quite remarkably, we find that the RILM prediction follows a dipole-fit at 
low-momenta, but falls-off faster at large momentum transfer, in agreement
with what is observed in the recoil polarization measurements.
Notice that this property of the form factor could not be 
understood at the level of the interaction of partons
with a single-instanton (Fig. \ref{Sachs}~B).

From the low-momentum points we can extract the proton 
charge radius, which  falls slightly short of 
the experimental value: $\la~R_{E\,(RILM)}^2\ra=(0.76~\fm)^2$ (to be compared
with  $\la R_{E\,(\exp.)}^2\ra= (0.81~\fm)^2~ $).
The fact that we obtain a slightly small charge radius is not surprising. 
Indeed, on the one hand we recall that in the present calculation we have 
used quarks of mass of about 70~MeV, 
corresponding to a rather heavy nucleon ($M=1.15~\GeV$).
On the other hand, we have neglected fermionically disconnected graphs,
which encode some of the sea contribution (notice however
that some ``pion cloud'' contribution is present through the Z-graphs).

\begin{figure}
\includegraphics[scale=0.4,clip=]{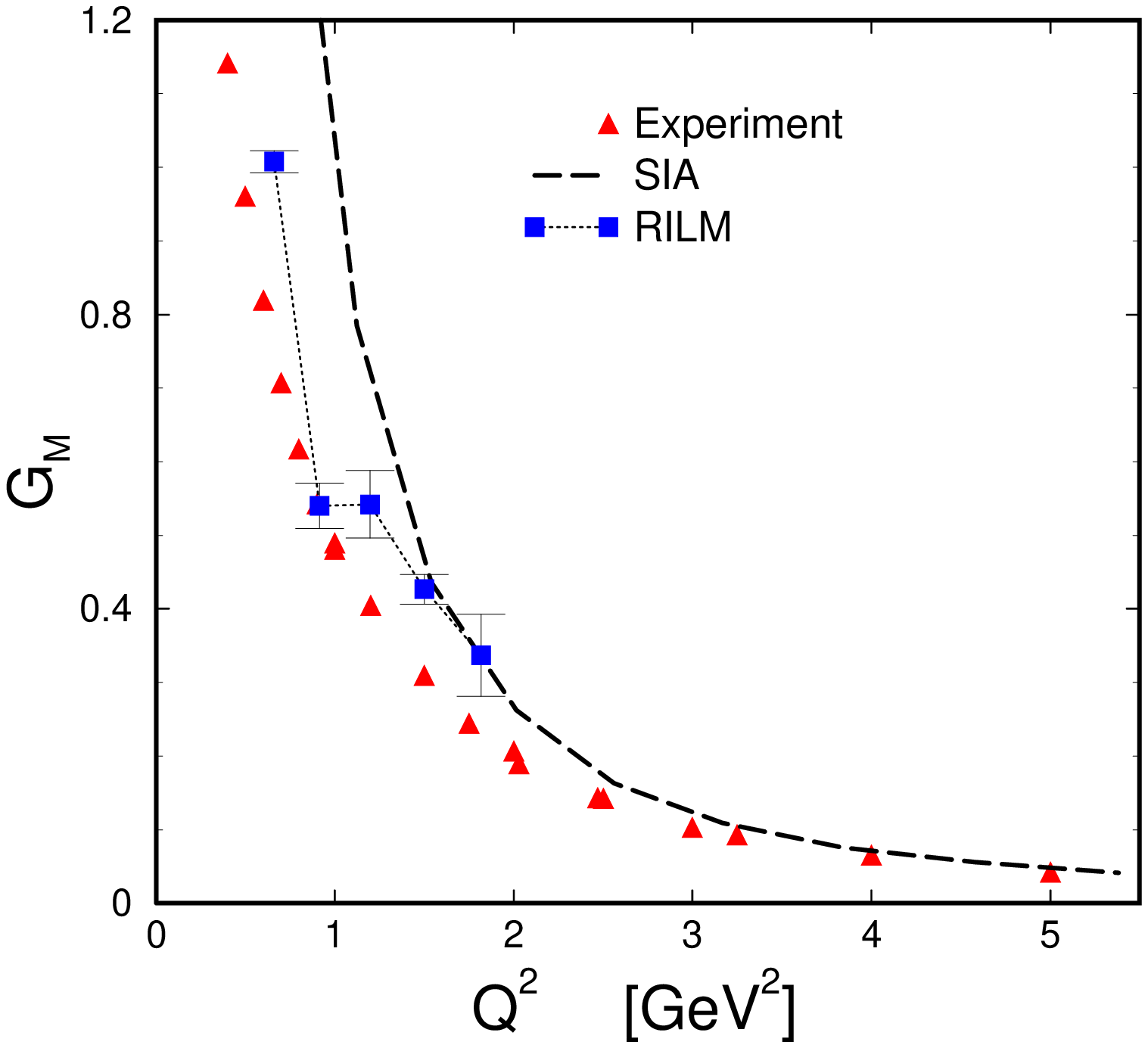}
\hspace*{1.5cm}
\includegraphics[scale=0.4,clip=]{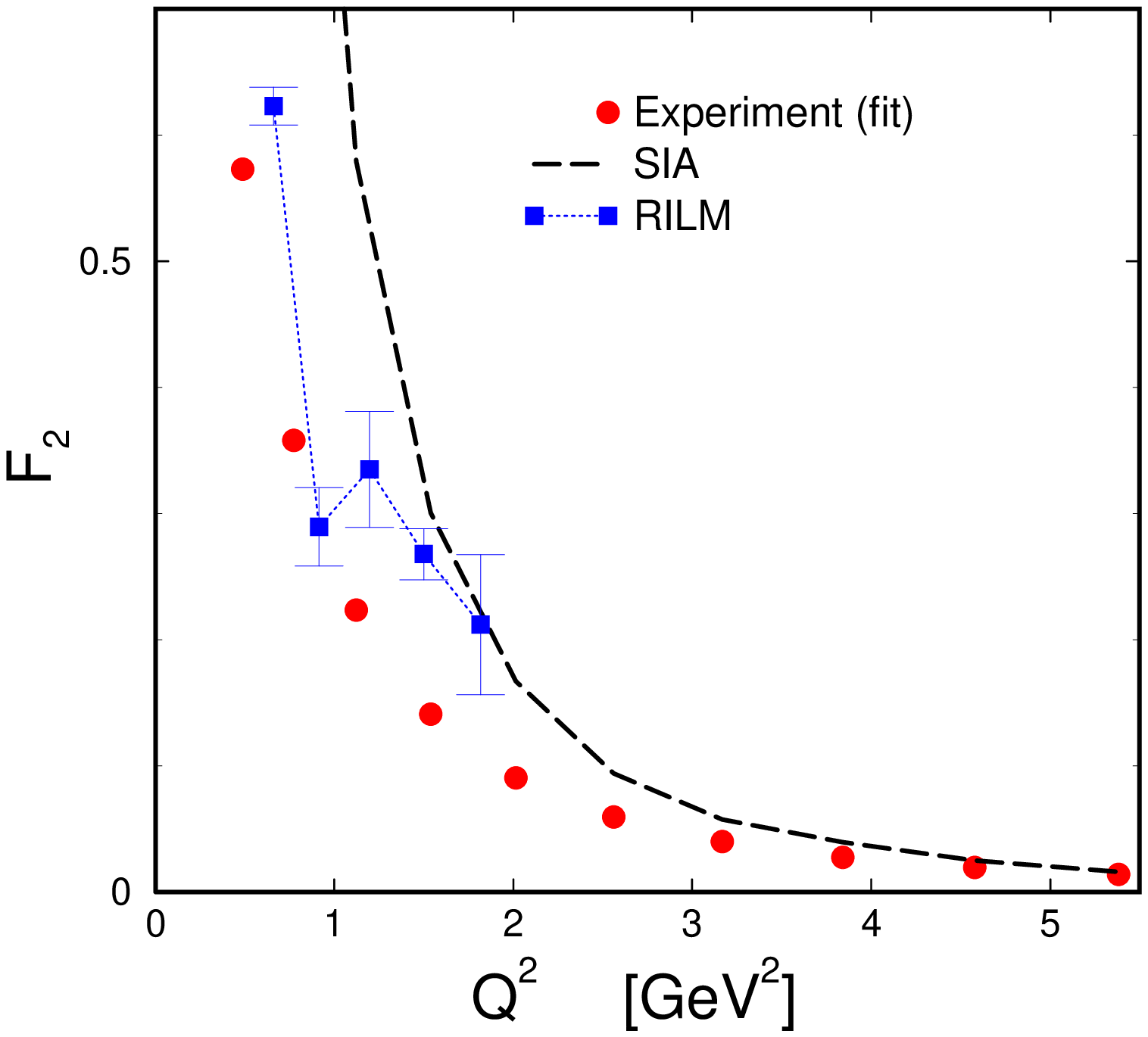}
{\large \qquad\qquad(A)\hspace*{7.5cm}(B)}
\caption{(Color online)$\mbox{}$\\ 
(A): Magnetic form factor of the proton in the ILM (squares) 
and from experiment (triangles).
The ILM curve has been obtained by combining the analytical 
SIA prediction for the  
Dirac form factor $F_1$ with the numerical RILM results for $G_E$.\\
(B): Pauli Form Factor of the proton
in the ILM. The circles are obtained from a fit of the
experimental data. Squares is the ILM prediction, obtained by combining
RILM results for $G_E$ with the SIA results for~$F_1$.}
\label{GMILM}
\end{figure}

In the previous section, we have shown that the proton Dirac form 
factor is completely saturated by the one-instanton contribution, already
at relatively low momenta ($Q\gtrsim 0.5~\GeV^2$).
We can use this result, to combine the
RILM result for $G_E(Q^2)$ and the SIA result for $F_1(Q^2)$ 
and obtain the Magnetic and  the Pauli form factor
of the proton in the ILM, for $Q\gtrsim 0.5~\GeV^2$.
The results for such form factors are reported in Fig. \ref{GMILM}.
Also in these two cases we see that the inclusion of many-instanton effects
is sufficient to explain the low-energy data.

\subsection{Electric Form Factor of the Neutron in the ILM}

\begin{figure}
\includegraphics[scale=0.4,clip=]{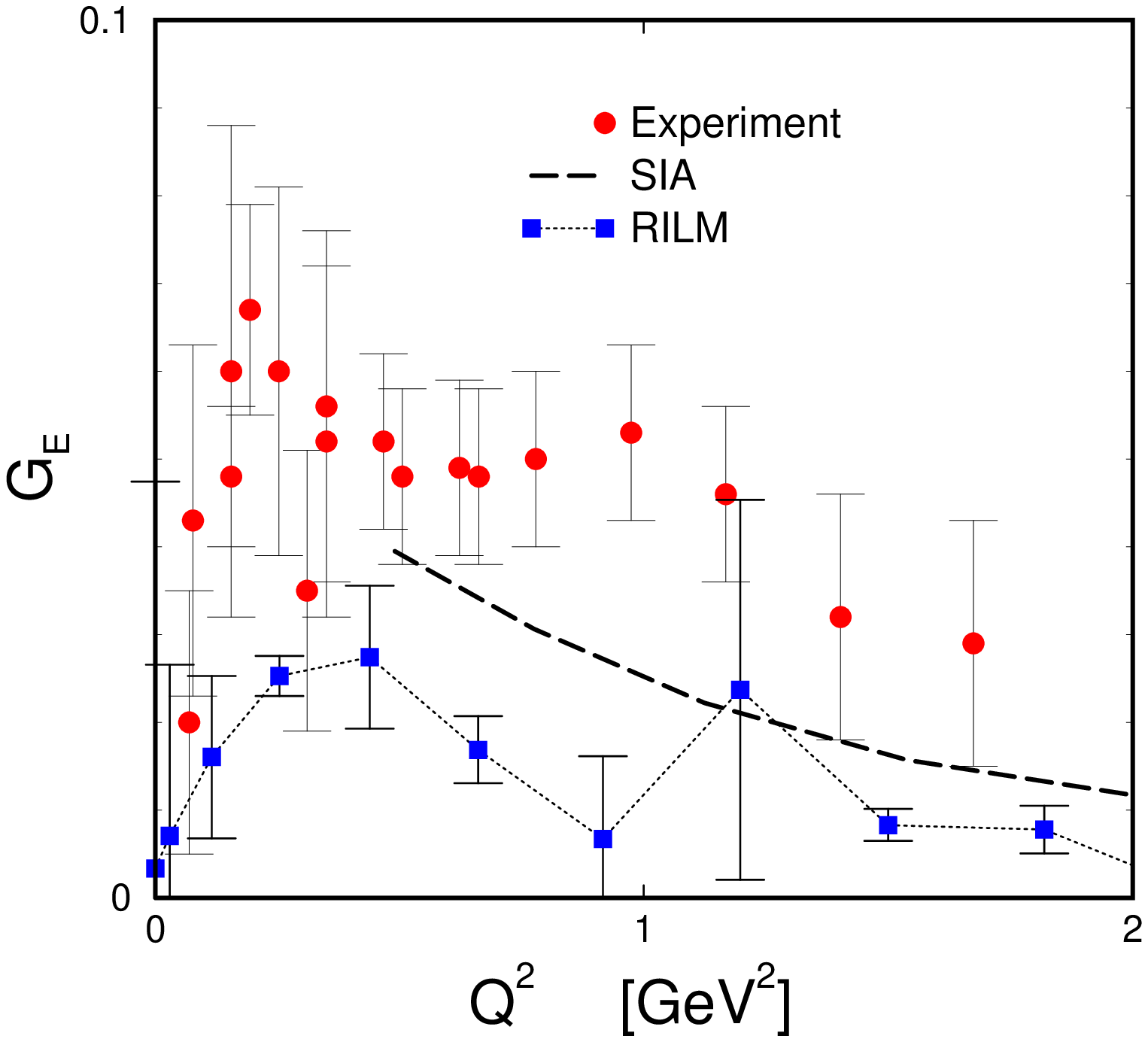}
\hspace*{1.5cm}
\caption{(Color online) $\mbox{}$\\
Electric form factor of the neutron in the RILM 
 and from experiment \cite{NGEexp}. 
Circles are experimental data, squares are RILM point,
while the dashed line is the SIA curve.}
\label{GEnfig}
\end{figure}

The  results from the 
electric form factor of the neutron are shown  in Fig. \ref{GEnfig}.
In this case, the agreement with experiment is somewhat
worse than the corresponding
results for the proton electric form factor. Our 
theoretical prediction  undershoots experimental data by a factor 2 or so.
We believe that  this discrepancy is mainly due
to the absence of disconnected graphs.
Clearly, the relative contribution
of such  SU(3) breaking effects are much more important in the case 
of the neutron, which has a very small electric form factor compared to
the proton. This hypothesis is supported by the fact that in 
\cite{Lattice1}, the disconnected diagrams
were calculated in lattice QCD and found to give a contribution
of the order of $50\%$ to the form factor.
A systematic study of the sea contribution coming from
disconnected graphs to several low-energy
observables is currently in progress \cite{PJD}.

\section{Conclusions}
\label{conclusions}

The present study was motivated by the observation that recent
JLAB data show that electro-magnetic form factors 
are very sensitive to some short-distance non-perturbative dynamics.
Instantons are known to play the leading role in
the spontaneous breaking of chiral symmetry and in the saturation of the
chiral anomaly, i.e. in two very important non-perturbative phenomena
which occur at the GeV scale.
In a previous analysis we had showed that instantons saturate the
the pion charged form factor and, at the same time, 
explain why the perturbative
regime is reached much earlier in the $\gamma\,\gamma^*\to~\pi_0$ transition 
from factor.
In this work, we have asked whether they can also
explain some existing puzzles concerning the nucleon form factors.
 
We have found that large momentum transfer data 
of Sachs as well as Pauli and Dirac form factors can be already reproduced
 by accounting for the scattering of the partons on a single-instanton.
These calculations have been carried-out in the SIA.
On the other hand,  form factors at low-momenta  cannot be calculated
 in the SIA because, in such a kinematic regime, 
many-instanton effects are very important. The only exception is the proton
Dirac form factor, which is saturated by one instanton, already
at relatively low momentum transfer ($Q^2\sim 0.5~\GeV^2$).

We have evaluated numerically the electric form factors in the full instanton
vacuum, i.e. to all orders in the 't~Hooft interaction, using the RILM.
In the case of the proton, we have found that RILM predictions are 
consistent with SIA calculations at large
momentum transfer and quantitatively reproduce the available body
of experimental data.
In particular, we have shown that in the ILM the electric form factor
follows a dipole fit at low momenta, 
but falls-off faster at large momenta, in quantitative agreement 
with the recent JLAB results. 
On the other hand, the electric form factor of the neutron
seems to be rather sensitive to fermionically disconnected graphs and
SU(3) breaking effects, which
have been neglected in the present approach.

We have combined our SIA result for the proton Dirac form factor, with our
numerical RILM results for the electric form factors and obtained predictions
for the magnetic and Pauli form factors.
As in the previous cases, we have found very good agreement with 
experiment for all proton form factors.
In the future we are planning to use the framework developed in this work
to investigate the role of the pion cloud in low-energy observables.
\acknowledgments
I would like thank E.V.~Shuryak and A.~Schwenk for their help and support.
The code for calculating vacuum expectation values in the 
instanton vacuum has been kindly made available
by  E.V.~Shuryak  and T.~Sch\"afer.
I also acknowledge interesting 
discussions with D.~Guadagnoli, 
F.~Iachello, G.A.~Miller, J.W.~Negele,  S.~Simula and W.~Weise.
Part of this work was performed while visiting
the Physics Department of the ``Universita' di Roma 3'', which I 
thank for the kind hospitality.

\newpage
\section*{APPENDIX: 
ANALYTIC RESULTS IN THE SIA}

Every where, we choose ${\bf q}$ pointing along the $\hat{1}$ 
direction:  ${\bf q}=(q,0,0)$. Let us define the following functions:
\be
\xi^-(t):= \sqrt{(t-z_4)^2+\bar{\rho}^2}\nonumber\\
\xi^+(t):= \sqrt{(t+z_4)^2+\bar{\rho}^2}\nonumber\\
\xi^0(t):= \sqrt{t^2+\bar{\rho}^2}
\ee

\subsection{Two-point function}
The SIA result for the two-point function defined in (\ref{G2}) is:
\be
\label{G2analytic}
{\bf G2}(t,{\bf q})=\frac{32\,\bar{n}\,\rho^4}{m^{*\,2}\,\pi^6}
\int_{0}^{\infty} d |{\bf k}| ~d |{\bf l}|
~ d |{\bf m}| ~|{\bf m}|^2\,|{\bf k}|^2\,|{\bf l}|^2
\int_{-1}^{1}  d c_k~ d c_l~ d c_m
\int_{-\infty}^{\infty} d z_4 
~\left( A + B  \right),\nonumber
\ee
where
\be
A:=
2\,K_0(|{\bf k}|\,\xi^-(t/2))\,K_0(|{\bf m}|\,\xi^+(t/2))\,
K_0(\chi_m^-\,\xi^+(t/2))\,K_0(\chi_k^-\,\xi^-(t/2))
\,e^{-t\,\chi_q^+},\nonumber\\
B:=
\,K_0(|{\bf k}|\,\xi^-(t/2))\,K_0(|{\bf m}|\,\xi^+(t/2))
\,K_0(\chi_m^+\,\xi^+(t/2))\,K_0(\chi_k^+\,\xi^-(t/2))
\,e^{-t\,\chi_q^-},\nonumber
\ee
and
\be
\chi_Q^+:=\sqrt{|{\bf l}|^2+|{\bf q}|^2+2\,
|{\bf q}|\,|{\bf l}|\,c_l}\nonumber\\
\chi_Q^-:=\sqrt{|{\bf l}|^2+|{\bf q}|^2-2\,
|{\bf q}|\,|{\bf l}|\,c_l}\nonumber\\
\chi_m^+:=\sqrt{|{\bf l}|^2+|{\bf m}^2+2\,|{\bf m}\,|{\bf l}|\,c_m}\nonumber\\
\chi_m^-:=\sqrt{|{\bf l}|^2+|{\bf m}^2-2\,|{\bf m}\,|{\bf l}|\,c_m}\nonumber\\
\chi_k^+:=\sqrt{|{\bf l}|^2+|{\bf k}|^2+
2\,|{\bf k}|\,|{\bf l}|\,c_k}\nonumber\\
\chi_k^-:=\sqrt{|{\bf l}|^2+|{\bf k}|^2-2\,|{\bf k}|\,|{\bf l}|\,c_k}\nonumber.
\ee

\subsection{Magnetic Three-point Function}

The SIA result for the proton (neutron) magnetic three-point 
function defined in (\ref{G3M}) is:
\be
{\bf G3}^{p(n)}_M(t,{\bf q})=
\frac{\bar{n}\,\bar{\rho}^4}{m^{*\,2}}\,
\int \frac{d^3 {\bf k}}{(2\,\pi)^3}
\int \frac{d^3 {\bf l}}{(2\,\pi)^3}
\int \frac{d^3 {\bf m}}{(2\,\pi)^3}\,
\left[ \frac{2}{3}\,U^{M}(D)({\bf k},{\bf l},{\bf m})
-\frac{1}{3}\,D^{M}(U)({\bf k},{\bf l},{\bf m})
\right],\nonumber
\ee
where
\be
U^{M}({\bf k},{\bf l},{\bf m}):=
U^{M}_1+U^{M}_2+U^{M}_3+U^{M}_4,
\nonumber\ee
and
\be
D^{M}({\bf k},{\bf l},{\bf m}):=
D^{M}_{1}+D^{M}_2\nonumber.
\ee

The functions $U^{M}_{1-4}$ correspond to sets of diagrams
in which the virtual photon is absorbed by a $u$-quark. 
They are defined as follows:
\begin{itemize}
\item
$U^M_1:=u^M_{1a}+u^M_{1b}+u^M_{1c},$
\vspace{1cm}
\be
u^M_{1a}:= -256\,
\frac{m_3^2+m_1^2+ q m_1-m_2^2-|{\bf m}||{\bf m}+{\bf q}|}
{|{\bf m}||{\bf m}+{\bf q}|}
\,e^{-t\ (|{\bf m}+{\bf q}|+|{\bf m}|)}
\,K_0(|{\bf k}|\,\xi^-(t))\,\nonumber\\
\times\,
K_0(|{\bf l}|\,\xi^+(t))\,
K_0(|{\bf q}/2+{\bf m}+{\bf k}|\,\xi^-(t))\,
K_0(|{\bf q}/2+{\bf m}+{\bf l}|\,\xi^+(t))\nonumber.
\ee
\be
u^M_{1b}:=  +256\,
\frac{l_2\,k_2}{|{\bf l}||{\bf k}+{\bf q}|}\,
e^{-t (|{\bf k}+{\bf q}|+2|{\bf l}|)}\,
K_0(|{\bf k}|\,\xi^0(t))\,K_0(|{\bf q}/2+{\bf l}+{\bf k}|\,\xi^-(t))
\nonumber\\
\times\,K_0(|{\bf m}|\,\xi^+(t))\,
K_0(|{\bf q}/2+{\bf m}+{\bf l}|\,\xi^+(t))
\nonumber.\ee

\be
u^M_{1c}:= +256\,
\frac{-m_1\,l_1+m_2\,l_2-m_3\,l_3+|{\bf m}||{\bf l}|}{|{\bf m}||{\bf l}|}
\,e^{-t (|{\bf m}|+2|{\bf l}|)}
\,K_0(|{\bf m}+{\bf q}|\,\xi^0(t))
\nonumber\\
\times K_0(|{\bf k}|\,\xi^-(t))\,
K_0(|{\bf q}/2-{\bf l}-{\bf k}|\,\xi^-(t))\,
K_0(|{\bf q}/2+{\bf m}+{\bf l}|\,\xi^+(t))
\nonumber\ee

\item $U^M_2:=u^M_{2a}+u^M_{2b},$
\vspace{1cm}
\be
u^M_{2a}:= -1024
\frac{m_2(l_2 + m_2)}{|{\bf q}/2+{\bf m}+{\bf l}||{\bf m}|}
\,e^{-t (2|{\bf q}/2+{\bf m}+{\bf l}|+|{\bf m}|)}\,
K_0(|{\bf q}-{\bf k}+{\bf l}+{\bf m}|\,\xi^-(t))
\nonumber\\
\times\,
K_0(|{\bf m+q}|\,\xi^0(t))\,
K_0(|{\bf k}|\,\xi^-(t))\,
K_0(|{\bf l}|\,\xi^+(t))
\nonumber\ee
\be
u^M_{2b}:= -1024\, 
\frac{(l_2-k_2+m_2)\,(m_2+l_2)}
{|{\bf q}/2+{\bf m}+{\bf l}||{\bf q}-{\bf k}+{\bf m}+{\bf l}|}
e^{-t (2|{\bf q}/2+{\bf m}+{\bf l}|+|{\bf m} + {\bf q} - {\bf k} +{\bf l}|)}
\,K_0(|{\bf k}|\,\xi^-(t))
\nonumber\\
\times\,K_0(|{\bf l}+{\bf m}-{\bf k}|\,\xi^0(t))\,
K_0(|{\bf l}|\,\xi^+(t))\,
K_0(|{\bf m}|\,\xi^+(t))
\nonumber\ee
\item 
$ U^M_3:=u^M_{3a}+u^M_{3b}+u^M_{3\,c}$,
\vspace{1cm}
\be
u^M_{3a}:= +\frac{128}{|{\bf m}+{\bf l} - {\bf k} + {\bf q}|
|{\bf m}+{\bf q}/2+{\bf l}|}
(-2 l_1 k_1 + 2 l_1^2 + 3 q m_1 - 2 m_1 k_1 + 4 m_1 l_1
+2 m_1^2 
\nonumber\\
 -  q k_1 + 2 q l_1  + 3 q l_1- 2 l_3 k_3  +2 l_3^2 -2 m_3 k_3 + 4 m_3 l_3 
+q^2 -2 l_2^2  -2 m_2^2 
\nonumber\\
- 4 m_2 l_2 + 2 m_2 k_2 +2 l_2 k_2 + 2 m_3^2+
2\,|{\bf q}/2+{\bf m}+{\bf l}||{\bf q}+{\bf m}+ {\bf l}-{\bf k}|)\,\nonumber\\
\times\,e^{-t (2|{\bf q}/2+{\bf m}+{\bf l}|+|{\bf q}-{\bf k}+{\bf l}+{\bf m}|)}
K_0(|{\bf m}|\,\xi^+(t))
K_0(|{\bf l}|\,\xi^+(t))\,
K_0(|{\bf k}|\,\xi^-(t))\,
K_0(|{\bf m}+{\bf l}-{\bf k}|\,\xi^0(t))\nonumber
\\
\nonumber\ee
\be
u^M_{3b}:=  -256\,
\frac{(l_2 + m_2)\,m_2}{|{\bf m}||{\bf q}/2+{\bf m}+{\bf l}|}
\,e^{-t\,(2|{\bf m}+{\bf l}+{\bf q}/2|+|{\bf m}|)}\,
K_0(|{\bf k}|\,\xi^-(t))\,
K_0(|{\bf l}|\,\xi^+(t))\,\nonumber\\
K_0(|{\bf q}+{\bf m}|\,\xi^0(t))
K_0(|{\bf q}-{\bf k}+{\bf m}+{\bf l}|\,\xi^-(t))
\nonumber\ee
\be
u^M_{3c}:=  -256\,
\frac{m_1^2-m_2^2+m_3^2 + q\,m_1-|{\bf m}|
|{\bf m}+{\bf q}|}{|{\bf m}||{\bf m}+{\bf q}|}
e^{-t (2|{\bf m} + {\bf q}|+|{\bf m}|)}
K_0(|{\bf l}|\,\xi^+(t))\,\nonumber\\
K_0(|{\bf q}/2+{\bf m}+{\bf l}|\,\xi^+(t))\,
K_0(|{\bf k}|\,\xi^-(t))\,
K_0(|{\bf q}/2+{\bf m}+{\bf k}|\,\xi^-(t))
\nonumber\ee
\item $U^M_4:=u^M_{4a},$
\vspace{1cm}
\be
u^M_{4a}:=  -1024\,
\frac{(m_1^2-m_2^2+m_3^2+ 
q\,m_1-|{\bf m}||{\bf m}+{\bf q}|)\,m_2}{|{\bf m}||{\bf q}+{\bf m}|}
\,e^{-t\,(|{\bf m}+{\bf q}|+|{\bf m}|)}
K_0(|{\bf k}|\,\xi^-(t))\,\nonumber\\
K_0(|{\bf q}/2+{\bf m}+{\bf k}|\,\xi^-(t))\,
K_0(|{\bf q}/2+{\bf m}+{\bf l}|\,\xi^+(t))
K_0(|{\bf l}|\,\xi^+(t))
\nonumber\ee
\end{itemize}
The functions $D^M_{1-2}$ correspond to sets of diagrams
in which the photon is absorbed by a $d$-quark. They are defined as follows:
\begin{itemize}
\item
$D^M_1:= d^M_{1a}+d^M_{1b}$
\vspace{1cm}
\be
d^M_{1a}:= -1024\,
\frac{m_2\,(m_2+l_2)}{|{\bf m}||{\bf m}+{\bf q}/2+{\bf l}|}\,
e^{-t (|{\bf m}| + 2\,|{\bf q}/2+{\bf l}+{\bf m}|)}
K_0(|{\bf m}+{\bf q}|\,\xi^0(t))\nonumber\\
K_0(|{\bf l}|\,\xi^+(t))\,
K_0(|{\bf k}|\,\xi^-(t))\,
K_0(|{\bf q}-{\bf k}+{\bf m}+{\bf l}|\,\xi^-(t))
\nonumber\ee
\be
d^M_{1b}:= +1024\,
\frac{(m_2+l_2)\,(k_2-l_2-m_2)}
{|{\bf m}+{\bf q}/2+{\bf l}|+|{\bf q}+{\bf l}+{\bf m}-{\bf k}|}
\,e^{-t (2|{\bf m}+{\bf q}/2+{\bf l}|+|{\bf q}+{\bf l}+{\bf m}-{\bf k}|)}
\nonumber\\
K_0(|{\bf m}+{\bf l}-{\bf k}|\,\xi^0(t))\,
K_0(|{\bf m}|\,\xi^+(t))\,
K_0(|{\bf k}|\,\xi^-(t))\,
K_0(|{\bf l}|,\xi^+(t))
\nonumber\ee
\item $D^M_2:=d^M_{2a}+d^M_{2b}+d^M_{2c}+d^M_{2d}$,
\vspace{1cm}
\be
d^M_{2a}:= +256\, 
\frac{{\bf m}\,{\bf l} - |{\bf l}||{\bf m}|}{|{\bf m}|\,|{\bf l}|}
\,e^{-t (|{\bf m}|+2|{\bf l}|)}
K_0(|{\bf m}+{\bf q}|\,\xi^0(t))\,
K_0(|{\bf k}|\,\xi^+(t))\,\nonumber\\
K_0(|{\bf q}/2+{\bf m}+{\bf l}|\,\xi^+(t))\,
K_0(|{\bf q}/2-{\bf k}-{\bf l}|\,\xi^-(t))
\nonumber\ee
\be
d^M_{2b}:= -256\,
\frac{m_2\,(m_2+l_2)}{|{\bf m}+{\bf q}/2+{\bf l}||{\bf m}|}
\,e^{-t (2|{\bf m}+{\bf q}/2+{\bf l}|+|{\bf m}|)}
K_0(|{\bf m}+{\bf l}-{\bf k}+{\bf q}|\,\xi^-(t))\nonumber\\
K_0(|{\bf m}+{\bf q}|\,\xi^0(t))\,
K_0(|{\bf k}|\,\xi^-(t))\,
K_0(|{\bf l}|,\xi^+(t))
\nonumber\ee
\be
d^M_{2c}:= +256\,
\frac{l_2\,k_2}{|{\bf k}+{\bf q}||{\bf l}|}\,
e^{-t (|{\bf q}+{\bf k}|+2|{\bf l}|)}
K_0(|{\bf l}+{\bf k}+{\bf q}/2|\,\xi^-(t))\,
K_0(|{\bf m}|\,\xi^+(t))\,\nonumber\\
K_0(|{\bf k}|\,\xi^0(t))\,
K_0(|{\bf q}/2+{\bf m}+{\bf l}|,\xi^+(t))
\nonumber\ee
\be
d^M_{2d}:= -\frac{256}{2|{\bf m}+{\bf l}-{\bf k}+{\bf q}|
|{\bf q}/2+m+{\bf l}|} \,
e^{-t (|{\bf m}+{\bf l}-{\bf k}+{\bf q}|+2|{\bf l}+{\bf q}/2+{\bf m}|)}
\,K_0(|{\bf l}|\,\xi^+(t))\,
\nonumber\\
\times\,(2 m_1^2+3 m_1 q-q k_1-2 m_1 k_1-2 l_1 k_1 -2 m_3 k_3-2 l_3 k_3+q^2+
\nonumber\\ 
3 l_1 q+ 4 l_1 m_1+ 2 l_1^2+ 2 m_3^2 +4 l_3 m_3+ 2 l_3^2+ 4 l_2 m_2
\nonumber\\
-2 m_2 k_2-2 l_2 k_2 + 2 l_2^2 + 2 m_2^2+
2|{\bf q}/2+ {\bf m} + {\bf l}|\,|{\bf q}+{\bf m}+{\bf l}-{\bf k}|)
\nonumber\\
\times\,K_0(|{\bf m}-{\bf k}+{\bf l}|\,\xi^0(t))\,K_0(|{\bf k}|\,\xi^-(t))\,
K_0(|{\bf m}|,\xi^+(t))\nonumber\\
\nonumber\ee
\end{itemize}

\subsection{Electric Three-point Function}

The proton (neutron) electric three-point function reads:

\be
{\bf G3}^{p(n)}_E(t,{\bf q})=
\frac{\bar{n}\,\bar{\rho}^4}{m^{*\,2}}\,
\int \frac{d^3 {\bf k}}{(2\,\pi)^3}
\int \frac{d^3 {\bf l}}{(2\,\pi)^3}
\int \frac{d^3 {\bf m}}{(2\,\pi)^3}\,
\left[ \frac{2}{3}\,U^{E}(D)({\bf k},{\bf l},{\bf m})
-\frac{1}{3}\,D^{E}(U)({\bf k},{\bf l},{\bf m})
\right]\nonumber
\\
\nonumber\ee
where,
\be
U^{E}({\bf k},{\bf l},{\bf m}):=
U^{E}_1+U^{E}_2+U^{E}_3+U^{E}_4,
\ee
and
\be
D^{E}({\bf k},{\bf l},{\bf m}):=
D^{E}_{1}+D^{E}_2.
\nonumber\ee
As in the case of the magnetic three-point function, 
$U^{E}_{1-4}$ correspond to sets of diagrams
in which the photon is absorbed by a $u$-quark. They are defined as follows:

\begin{itemize}
\item
$U^E_1:=u^E_{1a}+u^E_{1b}+u^E_{1c},$
\vspace{1cm}
\be
u^E_{1a}:= -256\,
\frac{{\bf m}^2 + q m_1 + |{\bf m}||{\bf m}+{\bf q}|}
{|{\bf m}||{\bf m}+{\bf q}|}
\,e^{-t\ (|{\bf m}+{\bf q}|+|{\bf m}|)}
\,K_0(|{\bf k}|\,\xi^-(t))\,\nonumber\\
\times\,
K_0(|{\bf l}|\,\xi^+(t))\,
K_0(|{\bf q}/2+{\bf m}+{\bf k}|\,\xi^-(t))\,
K_0(|{\bf q}/2+{\bf m}+{\bf l}|\,\xi^+(t))\nonumber
\ee
\be
u^E_{1b}:=  -256\, e^{-t (|{\bf k}+{\bf q}|+2|{\bf l}|)}\,
K_0(|{\bf k}|\,\xi^0(t))\,K_0(|{\bf q}/2+{\bf l}+{\bf k}|\,\xi^-(t))
\nonumber\\
\times\,K_0(|{\bf m}|\,\xi^+(t))\,
K_0(|{\bf q}/2+{\bf m}+{\bf l}|\,\xi^+(t))
\nonumber\ee

\be
u^E_{1c}:= -256\,
\frac{{\bf l} {\bf m} + |{\bf m}||{\bf l}|}{|{\bf m}||{\bf l}|}
\,e^{-t (|{\bf m}|+2|{\bf l}|)}
\,K_0(|{\bf m}+{\bf q}|\,\xi^0(t))
\nonumber\\
\times K_0(|{\bf k}|\,\xi^-(t))\,
K_0(|{\bf q}/2-{\bf l}-{\bf k}|\,\xi^-(t))\,
K_0(|{\bf q}/2+{\bf m}+{\bf l}|\,\xi^+(t))
\nonumber\ee

\item $U^E_2:=u^E_{2a}+u^E_{2b},$
\vspace{1cm}
\be
u^E_{2a}:= -1024\,e^{-t (2|{\bf q}/2+{\bf m}+{\bf l}|+|{\bf m}|)}\,
K_0(|{\bf q}-{\bf k}+{\bf l}+{\bf m}|\,\xi^-(t))
\nonumber\\
\times\,
K_0(|{\bf m+q}|\,\xi^0(t))\,
K_0(|{\bf k}|\,\xi^-(t))\,
K_0(|{\bf l}|\,\xi^+(t))
\nonumber\ee
\be
u^E_{2b}:= -1024\, 
e^{-t (2|{\bf q}/2+{\bf m}+{\bf l}|+|{\bf m} + {\bf q} - {\bf k} +{\bf l}|)}
\,K_0(|{\bf k}|\,\xi^-(t))
\nonumber\\
\times\,K_0(|{\bf l}+{\bf m}-{\bf k}|\,\xi^0(t))\,
K_0(|{\bf l}|\,\xi^+(t))\,
K_0(|{\bf m}|\,\xi^+(t))
\nonumber\ee
\item 
$ U^E_3:=u^E_{3a}+u^E_{3b}+u^E_{3\,c}$,
\vspace{1cm}
\be
u^E_{3a}:= +\frac{128}{|{\bf m}+{\bf l} - {\bf k} + {\bf q}|
|{\bf m}+{\bf q}/2+{\bf l}|}
\,e^{-t (2|{\bf q}/2+{\bf m}+{\bf l}|+|{\bf q}-{\bf k}+{\bf l}+{\bf m}|)}
\nonumber\\
\times\,( 2 {\bf m}^2 + 2 {\bf l}^2  +4 {\bf m} {\bf l} -2 {\bf l} {\bf k}
-2 {\bf m} {\bf k} + 3 q m_1 + 3 q l_1 
+q^2  - q k_1 
\nonumber\\
+ 
2\,|{\bf q}/2+{\bf m}+{\bf l}||{\bf q}+{\bf m}+ {\bf l}-{\bf k}|)\,
\,K_0(|{\bf m}|\,\xi^+(t))\nonumber\\
\times
K_0(|{\bf l}|\,\xi^+(t))\,
K_0(|{\bf k}|\,\xi^-(t))\,
K_0(|{\bf m}+{\bf l}-{\bf k}|\,\xi^0(t))\nonumber
\\
\nonumber\ee
\be
u^E_{3b}:=  -256\,
e^{-t\,(2|{\bf m}+{\bf l}+{\bf q}/2|+|{\bf m}|)}\,
K_0(|{\bf k}|\,\xi^-(t))\,
K_0(|{\bf l}|\,\xi^+(t))\,\nonumber\\
K_0(|{\bf q}+{\bf m}|\,\xi^0(t))
K_0(|{\bf q}-{\bf k}+{\bf m}+{\bf l}|\,\xi^-(t))
\nonumber\ee
\be
u^E_{3c}:=  -256\,
\frac{{\bf m}^2  + q\,m_1+|{\bf m}|
|{\bf m}+{\bf q}|}{|{\bf m}||{\bf m}+{\bf q}|}
e^{-t (2|{\bf m} + {\bf q}|+|{\bf m}|)}
K_0(|{\bf l}|\,\xi^+(t))\,\nonumber\\
K_0(|{\bf q}/2+{\bf m}+{\bf l}|\,\xi^+(t))\,
K_0(|{\bf k}|\,\xi^-(t))\,
K_0(|{\bf q}/2+{\bf m}+{\bf k}|\,\xi^-(t))
\nonumber\ee
\item $U^E_4:=u^E_{4a},$
\vspace{1cm}
\be
u^E_{4a}:=  -1024\,
\frac{{\bf m}^2 +  
q\,m_1+|{\bf m}||{\bf m}+{\bf q}|}{|{\bf m}||{\bf q}+{\bf m}|}
\,e^{-t\,(|{\bf m}+{\bf q}|+|{\bf m}|)}
K_0(|{\bf k}|\,\xi^-(t))\,\nonumber\\
K_0(|{\bf q}/2+{\bf m}+{\bf k}|\,\xi^-(t))\,
K_0(|{\bf q}/2+{\bf m}+{\bf l}|\,\xi^+(t))
K_0(|{\bf l}|\,\xi^+(t))
\nonumber\ee
\end{itemize}
The functions $D^E_{1-2}$ correspond to sets of diagrams
in which the photon is absorbed by a $d$-quark. They are defined as follows:

\begin{itemize}
\item
$D^E_1:= d^E_{1a}+d^E_{1b}$
\vspace{1cm}
\be
d^E_{1a}:=-1024\,
e^{-t (|{\bf m}| + 2\,|{\bf q}/2+{\bf l}+{\bf m}|)}
K_0(|{\bf m}+{\bf q}|\,\xi^0(t))\,K_0(|{\bf l}|\,\xi^+(t))
\nonumber\\
K_0(|{\bf k}|\,\xi^-(t))\,
K_0(|{\bf q}-{\bf k}+{\bf m}+{\bf l}|\,\xi^-(t))
\nonumber\ee
\be
d^E_{1b}:= -1024\,
\,e^{-t (2|{\bf m}+{\bf q}/2+{\bf l}|+|{\bf q}+{\bf l}+{\bf m}-{\bf k}|)}
\,K_0(|{\bf m}+{\bf l}-{\bf k}|\,\xi^0(t))
\nonumber\\
K_0(|{\bf m}|\,\xi^+(t))\,
K_0(|{\bf k}|\,\xi^-(t))\,
K_0(|{\bf l}|\,\xi^+(t))
\nonumber\ee
\item $D^E_2:=d^E_{2a}+d^E_{2b}+d^E_{2c}+d^E_{2d}$,
\vspace{1cm}
\be
d^E_{2a}:= +256\, 
\frac{{\bf m}\,{\bf l} - |{\bf l}||{\bf m}|}{|{\bf m}|\,|{\bf l}|}
\,e^{-t (|{\bf m}|+2|{\bf l}|)}
K_0(|{\bf m}+{\bf q}|\,\xi^0(t))\,
K_0(|{\bf k}|\,\xi^+(t))\,\nonumber\\
K_0(|{\bf q}/2+{\bf m}+{\bf l}|\,\xi^+(t))\,
K_0(|{\bf q}/2-{\bf k}-{\bf l}|\,\xi^-(t))
\nonumber\ee
\be
d^E_{2b}:= -256\,
\,e^{-t (2|{\bf m}+{\bf q}/2+{\bf l}|+|{\bf m}|)}
K_0(|{\bf m}+{\bf l}-{\bf k}+{\bf q}|\,\xi^-(t))\nonumber\\
K_0(|{\bf m}+{\bf q}|\,\xi^0(t))\,
K_0(|{\bf k}|\,\xi^-(t))\,
K_0(|{\bf l}|\,\xi^+(t))
\nonumber\ee
\be
d^E_{2c}:= -256\,
e^{-t (|{\bf q}+{\bf k}|+2|{\bf l}|)}
K_0(|{\bf l}+{\bf k}+{\bf q}/2|\,\xi^-(t))\,
K_0(|{\bf m}|\,\xi^+(t))\,\nonumber\\
K_0(|{\bf k}|\,\xi^0(t))\,
K_0(|{\bf q}/2+{\bf m}+{\bf l}|\,\xi^+(t))
\nonumber\ee
\be
d^E_{2d}:= \frac{-256}{2|{\bf m}+{\bf l}-{\bf k}+{\bf q}|
|{\bf q}/2+m+{\bf l}|} \,
e^{-t (|{\bf m}+{\bf l}-{\bf k}+{\bf q}|+2|{\bf l}+{\bf q}/2+{\bf m}|)}
\,K_0(|{\bf l}|\,\xi^+(t))\,
\nonumber\\
\times\,(2 m_1^2+3 m_1 q-q k_1-2 m_1 k_1-2 l_1 k_1 -2 m_3 k_3-2 l_3 k_3+q^2+
\nonumber\\ 
3 l_1 q+ 4 l_1 m_1+ 2 l_1^2+ 2 m_3^2 +4 l_3 m_3+ 2 l_3^2+ 4 l_2 m_2
\nonumber\\
-2 m_2 k_2-2 l_2 k_2 + 2 l_2^2 + 2 m_2^2+
2|{\bf q}/2+ {\bf m} + {\bf l}|\,|{\bf q}+{\bf m}+{\bf l}-{\bf k}|)
\nonumber\\
\times\,K_0(|{\bf m}-{\bf k}+{\bf l}|\,\xi^0(t))\,K_0(|{\bf k}|\,\xi^-(t))\,
K_0(|{\bf m}|\,\xi^+(t))\nonumber\\
\nonumber\ee
\end{itemize}


\begin{thebibliography}{99}

\bibitem{Fpi}
J. Volmer {\it et al.} [The Jefferson Laboratory $F_\pi$
Collaboration], Phys. Rev. Lett. \textbf{86} (2001) 1713.
%
%
\bibitem{JLAB1} M.K. Jones, \emph{et al.},
Phys. Rev. Lett. \textbf{84} (2000) 1398.
\bibitem{JLAB2} O. Gayou,  \emph{et al.}, Phys. Rev. Lett.
\textbf{88} (2002) 092301.
%
%
\bibitem{BF} S.J. Brodsky and G.R. Farrar,
Phys. Rev. Lett. \textbf{31} (1973) 1153.
\bibitem{Ji} A. Belitsky, X. Ji and F. Yuan,
Phys. Rev. Lett. \textbf{91} (2003) 092003.
\bibitem{chu94}
M.C. Chu, J.M. Grandy, S. Huang, and J.W. Negele,
Phys. Rev. \textbf{D49} (1994) 6039.
\bibitem{'thooft} 
G. 't~Hooft, Phys. Rev. Lett. \textbf{37} (1976) 8.\\
G. 't~Hooft, Phys. Rev. \textbf{D14} (1976) 3432.
\bibitem{dyakonov}
D. Diakonov, \emph{ Chiral symmetry
breaking by instantons}, Lectures given at the "Enrico Fermi"
school in Physics, Varenna, June 25-27 1995, hep-ph/9602375.
\bibitem{shuryakrev} T. Sch\"afer and E.V. Shuryak,
Rev. Mod. Phys. \textbf{70} (1998) 323.
\bibitem{Degrand} T.A.~DeGrand and A.~Hasenfratz,
Phys. Rev. \textbf{D64} (2001) 034512.
\bibitem{scalar} P.~Faccioli and T.A.~DeGrand, Phys. Rev. Lett. \textbf{91} 
(2003) 182001.  
\bibitem{shuryak82} E.V. Shuryak,
Nucl. Phys. \textbf{B214} (1982) 237.
\bibitem{incomplete}
F.~Iachello, A.D.~Jackson and A.~Lande,
Phys. Lett. \textbf{B43} (1973) 191. \\
B.Q.~Ma, D.~Qing and I.~Smidt, Phys. Rev. \textbf{C65} (2002) 035205.\\
G.A.~Miller, Phys. Rev. \textbf{C66} (2002) 032201.\\
C.V.~Christov, A.Z.~Gorzki, K.~Goeke and P.V.~Pobilitza,
Nucl. Phys. \textbf{A592} (1995) 513. 
\bibitem{forkel} H. Forkel and M. Nielsen, Phys. Lett.
\textbf{B345} (1995) 55. 
\bibitem{forkel2} V.V. Braguta and A.I. Onishchenko, hep-ph/0311146.
\bibitem{blotz} 
A. Blotz and E.V. Shuryak, Phys. Rev. \textbf{D55} (1997) 4055.
\bibitem{3ptILM} P. Faccioli and E.V. Shuryak,
Phys. Rev. \textbf{D65} (2002) 076002.
\bibitem{sia} P. Faccioli and E.V. Shuryak,
Phys. Rev. \textbf{D64} (2001) 114020.
\bibitem{mymasses} P. Faccioli,
Phys. Rev.  {\bf D65} (2002) 094014.
\bibitem{pionFF} P. Faccioli, A. Schwenk and E.V. Shuryak,
Phys. Rev. {\bf D67} (2003) 113009.
\bibitem{Dorokhov}
A.~E.~Dorokhov,
JETP Lett.\  {\bf 77}, 63 (2003)
[Pisma Zh.\ Eksp.\ Teor.\ Fiz.\  {\bf 77}, 68 (2003)]


\bibitem{nucleonGE} P. Faccioli, A. Schwenk and E.V. Shuryak, Phys. Lett.
\textbf{B549} (2002) 93.
\bibitem{brown78}
L.S. Brown, R.D. Carlitz, D.B. Craemer  and C. Lee,
Phys. Rev. \textbf{D17} (1978) 1583.
\bibitem{PGEGMexp}
L.E. Price, {\it et al.}
Phys.  Rev.  {\bf D4} (1971) 45.\\
P.~Bosted {\it et al.},
Phys. Rev. Lett  {\bf 68} (1992) 3841.\\
R.C.~Walker {\it et al.}, Phys. Rev. {\bf D49} (1994), 5671.
%
%
\bibitem{NGEexp} M. Meyerhoff \emph{et al.}, Phys. Lett. \textbf{B327}
(1994) 201.\\
J. Becker \emph{et al.}, Eur. Phys. J. \textbf{A6}
(1999) 329.\\
C. Herberg \emph{et al.}, Eur. Phys. J. \textbf{A5}
(1999) 131.\\
M. Orstick \emph{et al.},
Phys. Rev. Lett. \textbf{83} (1999) 276.\\
I. Passchier \emph{et al.},
Phys. Rev. Lett. \textbf{82} (1999) 4988.\\
D. Rohe \emph{et al.}, Phys. Rev. Lett. \textbf{83}
(1999) 4257.\\
H. Zhu \emph{et al.}, Phys. Rev. Lett. \textbf{87}
(2001) 081801.
\bibitem{Negele}
J.W. Negele, Nucl. Phys. Proc. Suppl. \textbf{73} (1999) 92.
\bibitem{PJD} P.~Faccioli, J.~Negele and D.~Renner, 
in preparation.
\bibitem{martinelli} G. Martinelli and C.T. Sachrajda,
Nucl. Phys. \textbf{B316} (1989) 355.
\bibitem{draper} W. Wilcox, T. Draper and K.F. Liu,
Phys. Rev. \textbf{D46} (1992) 1109.
\bibitem{corrbaryons} T. Sch\"afer, E.V. Shuryak and J.J.M. Verbaarschot,
Nucl. Phys. \textbf{B412} (1994) 143.
\bibitem{Lattice1} S.J. Dong, K.F. Liu and A.G. Williams,
Phys. Rev. \textbf{D58} (1998) 074504.
\end{thebibliography}
\end{document}